%% file: cl0024wfzs.tex
\documentclass{aa}
\usepackage{graphicx}
\usepackage{rotfloat}
\begin{document}
\title{A wide-field spectroscopic survey of the cluster of galaxies
  \object{Cl0024+1654}}  
\subtitle{I. The catalogue\thanks{
    Based on observations obtained with the Canada--France--Hawaii
    Telescope and the William--Herschel Telescope.}
  }

\author{Oliver Czoske\inst{1}
  \and Jean-Paul Kneib\inst{1}
  \and Genevi\`eve Soucail\inst{1}
  \and Terry J.\,Bridges\inst{2}
  \thanks{\emph{Present address:}
    Anglo--Australian Telescope, PO Box 296, Epping NSW 1710,
    Australia}
  \and Yannick Mellier\inst{3,5}
  \and Jean-Charles Cuillandre\inst{4,5}
  }

\offprints{O. Czoske, \email{oczoske@ast.obs-mip.fr}}

\institute{
  Observatoire Midi-Pyr\'en\'ees, UMR5572, 14 Av.\ Edouard Belin, 
  31400 Toulouse, France 
  \and Institute of Astronomy, Madingley Road, Cambridge, CB3 0HA, UK
  \and Institut d'Astrophysique de Paris, 98bis Bd. Arago, 
  75014 Paris, France
  \and CFHT Corporation, P.O. Box 1597, Kamuela, Hawaii 96743, USA
  \and Observatoire de Paris, DEMIRM, 61 Av.\ de l'Observatoire, 75014
  Paris, France 
  }

\date{Received too late, Accepted even later}

\authorrunning{O. Czoske et al.}
\titlerunning{Spectroscopic survey of Cl0024: I. The catalogue}

%%%%%%%%%%%%%%%% Abstract %%%%%%%%%%%%%%%%%%%%%%%%%%%%%%%%%%%%%%
\abstract{We present the catalogue of a wide-field CFHT/WHT
  spectroscopic survey of the lensing cluster  Cl0024+1654 at
  $z\!=\!0.395$. This catalogue contains 618 new spectra, of which
  581 have identified redshifts. Adding redshifts available from the
  literature, the final catalogue contains data for 687 objects with
  redshifts identified for 650 of them. 295 galaxies have redshifts in
  the range 
  $0.37\!<\!z\!<\!0.41$, i.~e.\ are cluster members or lie in the
  immediate neighbourhood of the cluster. The area covered by the survey
  is $21\!\times\!25\,\mathrm{arcmin}^2$ in size, corresponding to
  $4\times4.8\,h^{-2}\,\mathrm{Mpc}^2$ at the cluster redshift. The
  survey is 45\% complete down to $V\!=\!22$ over the whole field
  covered; within 3~arcmin of the cluster centre the completeness 
  exceeds 80\% at the same magnitude. A detailed completeness analysis
  is presented. The catalogue gives astrometric position, redshift, $V$
  magnitude and $V\!-\!I$ colour, as well as the equivalent
  widths for a number of lines. Apart from the cluster Cl0024+1654
  itself, three other structures are identified in redshift space: a
  group of galaxies at $z\!=\!0.38$, just in front of Cl0024+1654 and
  probably interacting with it, a close pair of groups of galaxies at
  $z\!\sim\!0.495$ and an overdensity of galaxies at $z\!\sim\!0.18$
  with no obvious centre. The spectroscopic catalogue will be used to
  trace the three-dimensional structure of the cluster Cl0024+1654 as
  well as study the physical properties of the galaxies in the cluster
  and in its environment.   
  }

\maketitle
%
%________________________________________________________________

%%%%%%%%%%%%%%% Introduction %%%%%%%%%%%%%%%%%%%%%%%%%%%%%%%%%
\section{Introduction}
\label{sec-introduction}

Clusters of galaxies are increasingly viewed not as simple isolated
and relaxed systems but as embedded in and connected to the general
large-scale structure in the Universe. This view of clusters in a
larger context has consequences for the interpretation of cluster
galaxy populations, cluster dynamics and mass estimates. 

Clusters of galaxies grow by continuously accreting galaxies and
groups of galaxies from the surrounding field, mostly along
filamentary structures. In the process, galaxies are transformed from
the predominantly blue, actively star-forming, spiral population
characteristic of the field to the red, passive and elliptical population
characteristic of the inner and denser regions of clusters  (Dressler
\cite{morphdensity}, Abraham et al.\ \cite{abraham}, Balogh et al.\
\cite{balogh}). Cluster galaxy populations evolve with redshift: rich 
clusters at high redshift contain a larger fraction of blue galaxies
than local ones (Butcher \& Oemler \cite{butcher-oemler1}, 
\cite{butcher-oemler2}; Cl0024+1654 is an example of a
``Butcher-Oemler'' cluster). The exact nature of the interaction of
infalling galaxies with the cluster environment (hot intra-cluster
medium, tidal gravitational field) and its influence on the morphology
of galaxies, their gas content and star-formation rates (as measured
by galaxy colours and spectral type) are as yet
ill-understood; hence the interest in investigating the ``infall
region'' beyond $\sim\!1\,h^{-1}\,\mathrm{Mpc}$ distance from the
cluster centre, where the transition from field to cluster galaxies is
taking place. The advent of new wide-field CCD mosaic cameras
available on a number of  large telescopes (e.~g.\ CFHT, CTIO, Subaru,
ESO2.2m) makes it possible to obtain photometric and morphological
information on 1 - 10 Mpc scales around the cluster centres. However,
wide-field investigation of clusters demands both imaging and
spectroscopic observations. Individual spectra of galaxies 
describe their spectral energy distribution and provide their redshift,
which is indispensable to produce a catalogue of cluster members
with radial velocity and information regarding their stellar content 
and star formation history. At present there is only a limited number
of clusters with more than $\sim\!200$ spectroscopically identified
member galaxies (e.~g.\ Abraham et al.\ \cite{abraham}), and
especially at high redshift ($z\!\ga\!0.2$) spectroscopically
well-studied clusters become very rare, mostly due 
to the fact that contamination by field galaxies increases rapidly
with redshift.   

The fact that clusters are not isolated systems also raises questions
concerning the traditional ways of estimating masses of clusters of
galaxies through different mass estimators: gravitational lensing
analyses, kinematical analyses from redshifts of cluster member
galaxies and X-ray observations.  

Gravitational lensing is
sensitive to the total integrated mass along the line-of-sight from
the observer to the lensed sources, weighted by the appropriate
combination of angular size distances between observer, lens and
source (e.~g.\ Schneider, Ehlers \& Falco \cite{SEF}, Bartelmann \&
Schneider \cite{bartelmann-schneider}). In the
presence of massive  structures other than the cluster along the line
of sight, the mass derived from gravitational lensing overestimates
the mass of the cluster proper. Large spectroscopic surveys provide
additional information needed to correctly interpret the lensing
analysis in this case. Metzler et al.\ (\cite{metzler}) investigate
the influence of the presence of filaments and groups of galaxies in
the vicinity of a cluster on weak lensing estimates of the cluster
mass and find that significant overestimates (up to a factor 1.5 to 2)
are possible and even likely. Similar investigations with comparable
results were conducted by Cen et al.\ (\cite{cen}) and Reblinsky et
al.\ (\cite{reblinsky}).  

A similar bias should be expected to affect 
measurements of velocity dispersions: if foreground or background
groups of galaxies in the immediate neighbourhood of the cluster are
added into the redshift histogram, but are not 
resolved and recognized as separate entities, the velocity dispersion
of the cluster itself will be overestimated. Generally only 30
to 50 member galaxies  are used to estimate the line-of-sight velocity
dispersion (and virial cluster mass; see e.~g.\ the large
compilations of Girardi et al.\ \cite{girardi1} and Girardi \&
Mezzetti \cite{girardi2}). Furthermore, the measured redshifts are   
generally concentrated within a relatively small region within a
projected radius of $\sim\! 500\,h^{-1}\,\mathrm{kpc}$ of the cluster
centre. Whereas one can argue that these numbers might be sufficient
for relaxed clusters with regular spatial and velocity distributions,
the various derived estimates will contain large systematic errors if
unresolved substructures are present. What can be obtained from
redshift surveys is a galaxy number density weighted line-of-sight
velocity dispersion averaged along the line-of-sight.  With a
sufficiently large number of cluster member redshifts it is possible
to measure the variation of the line-of-sight velocity dispersion with
projected distance from the cluster centre (Carlberg et al.\
\cite{carlberg}), but determination of even 
more detailed information on the dynamical status (e.~g.\ velocity
anisotropy profile) of a cluster requires a forbiddingly large number
of redshifts (e.~g.\ Merritt \cite{merritt-coma}).   

%Since X-ray surface brightness of the hot IGM depends on
%the square of the electron density, X-ray observations of clusters
%should be largely unaffected by low-contrast foreground and background
%structures and thus probe the gravitational potential of the cluster
%itself in a relatively clean way (although X-ray mass estimates depend
%of course on equilibrium assumptions for the state of the IGM). {\bf
%  (Could throw  this paragraph out - OC)}

Clearly, combining information coming from gravitational lensing (weak
and strong), the X-ray emission from the hot intra-cluster gas, the
Sunyaev-Zeldovich effect and the galaxy velocity distribution is the
best way to arrive at secure mass estimates for clusters of
galaxies (see e.~g.\ Castander et al.\ \cite{castander}). This is now
possible for several 
cases. Of particular interest are clusters for  
which a significant discrepancy between different mass estimates is
found. One such cluster is the well-known lensing cluster
Cl0024+1654\footnote{This cluster is commonly called Cl0024+1654,
  although the common Internet databases, such as 
NED (\texttt{http://nedwww.ipac.caltech.edu}), list it as 
ZwCl0024.0+1652 as it originally appeared in Zwicky (\cite{zwicky}).}. 
About 100 redshifts of galaxies in this cluster were obtained by
Dressler \& Gunn (\cite{dressler1}) and Dressler et al.\ (\cite{dressler2}),
resulting in a 
velocity dispersion of $\sigma\!\simeq\!1300\,\mathrm{km\,s}^{-1}$,
which is consistent with mass estimates derived from the spectacular
arc system in the cluster centre (Kassiola, Kovner \& Fort \cite{KKF},
Smail et al.\ \cite{smail}, Tyson et al.\ \cite{tyson}, Broadhurst et
al.\ \cite{broadhurst}). Cl0024+1654 was among the first clusters in which a
coherent shear signal due to weak lensing was found (Bonnet et al.\ 
\cite{BMF}). A crude mass estimate from this analysis was consistent
with the strong lensing and kinematical estimates. In addition to the
signal due to Cl0024+1654 itself, Bonnet et al.\ (\cite{BMF}) also
found a coherent signal to the north-east of the cluster centre in an
area where no obvious galaxy overdensity could be seen. The X-ray 
luminosity of Cl0024+1654 on the other hand is unusually low for a
cluster of this velocity dispersion, and mass estimates from the X-ray
observations are a factor of two to three lower than the lensing and
kinematical estimates (Soucail et al.\ \cite{soucail}). 

In order to better understand the dynamics of Cl0024+1654 and how it
is embedded in the surrounding large-scale structure, we have
conducted a wide-field spectroscopic survey at the
Canada-France-Hawaii Telescope (CFHT) and the William Herschel
Telescope (WHT) from 1992 to 1996. In this paper we present the
catalogue of the galaxies observed for this survey.   
Section \ref{sec-observations}  summarizes the photometric and 
spectroscopic observations. The data reduction and analysis are
presented in Section \ref{sec-reduction}. Section \ref{sec-discussion}
discusses some interesting global results of the survey and describes
some structures found in the redshift distribution that are not
directly related to the cluster Cl0024+1654. A summary is given in
Section \ref{sec-conclusions}. A detailed analysis of the dynamics of
the cluster itself and its environment as well as  the spectral
properties of its member galaxies will be the subject of a forthcoming
paper (Czoske et al.\ \cite{paper2}). 

Throughout this paper we use a Hubble constant $H_0 =
100\,h^{-1}\,\mathrm{km}\,\mathrm{s}^{-1}\,\mathrm{Mpc}^{-1}$,  
$\Omega_{\rm M} = 1$ and ${\Omega_\Lambda} = 0$, which gives a
physical scale of $3.195\, h^{-1}\,\mathrm{kpc}\,\mathrm{arcsec}^{-1}$
at the cluster redshift.

\section{Observations}
\label{sec-observations}

%%%%%%%%%%%%%%%%%%% Imaging %%%%%%%%%%%%%%%%%%%%%%%%%%%%%%%%%
\subsection{Imaging}
\label{ssec-imaging}

%Three broad-band images were used for creating and analysing the
%spectroscopic survey presented in this paper. 

We will use photometric results from two broad-band images in the
analysis of the spectroscopic survey presented in this paper. 

%Two 10~min V-band images of the cluster field were obtained
%using the \textsc{EMMI} camera on the ESO New Technology Telescope
%(NTT) on 17/18 October 1993. Given that the seeing was rather poor at
%$1\farcs7$, pixels were binned using a blocking factor of 2 in both
%directions, resulting in a final pixel scale of $0\farcs7$. Nine  
%separate images were assembled into a $3\times3$ mosaic covering
%$25\times25\, \mathrm{arcmin}^2$. The corresponding photometric
%catalogue, created using FOCAS, was the primary source for candidate
%selection for the spectroscopic survey.

On 26 September 1995, we obtained an I-band image using the UH8k
camera (Luppino, Metzger \& Miyazaki \cite{UH8k}) on CFHT. 
The I-band image was reduced chip by chip, i.~e.\ the final
  result is in the form of eight individual images, one for each chip
  of the camera.
The final images were obtained as the mean of 10 exposures of 1200~s
each, using sigma clipping ($2.5\sigma$ above the mean level and
$4\sigma$ below) to reject hot pixels and cosmic ray hits,
and have very good seeing of $0\farcs7$ FWHM; however, the background
is marred by stray light, presumably due to the bright star 47~Psc
($V\!=\!5.1$, spectral type M3)   
% and 48~Psc ($V\!=\!6.1$, K5) 
at $\sim\!50\arcmin$ distance from the centre of Cl0024+1654, although
outside the field of view of the UH8k. The I-band photometric
catalogue was obtained from the individually stacked chips using the
SExtractor package (Bertin \& Arnouts \cite{sextractor}) with a
threshold of $1.5\,\sigma$ and a minimum detection area of 5 pixels.
The catalogue contains more than $4\times10^4$ objects over a field of
about $28\times28\,\mathrm{arcmin}^2$, the limiting magnitude is
$I\!\sim\!24$. Unless otherwise noted, we use total magnitudes as
given by SExtractor's \texttt{MAG\_BEST}. 
  The internal errors on
  the I-band magnitudes (as given by SExtractor) are smaller than 0.05
  for $I\!<\!22.7$ (0.01 for $I\!<\!20.7$). Note that due to the 
  chip-wise reduction of the I-band image, the gaps between the chips
  (of typical width 5-10~arcsec)
  were not filled in during stacking and the photometric catalogue
  contains no objects from these regions.

\begin{figure*}
%%% for draft: no image
% \resizebox{\hsize}{!}{\includegraphics[draft]{figures/cl0024.ps}}
%%% for astro-ph: binned images
%%  \resizebox{\hsize}{!}{\includegraphics{figures/smcl0024.ps}}
%%% for publication: full images
  \resizebox{\hsize}{!}{\includegraphics{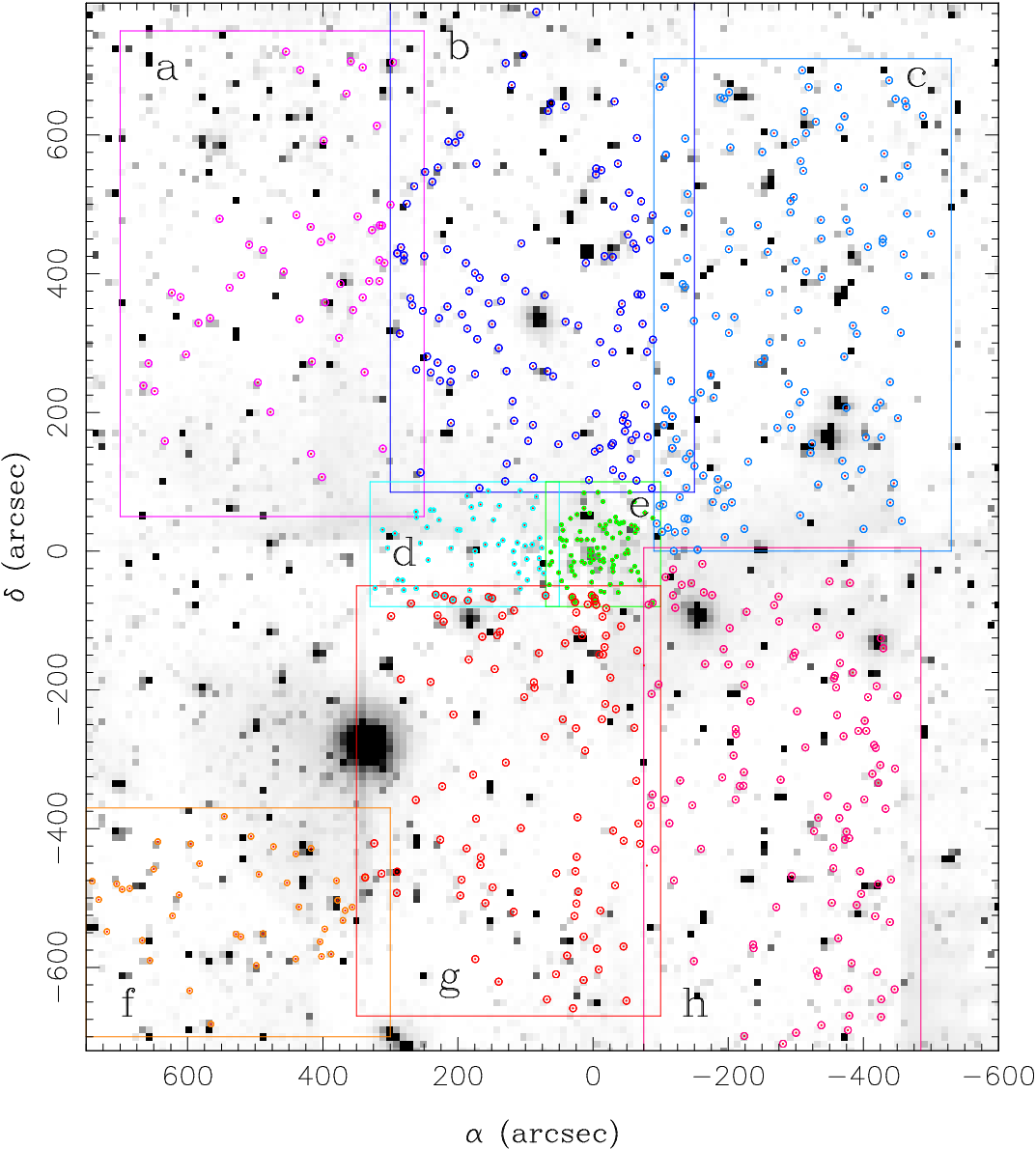}}
  \caption{$22\arcmin\times25\arcmin$ section of the CFH12k V-band
    image showing the distribution of the objects in our spectroscopic
    sample. Expanded views of the marked regions are shown in Figs.\ 
    \ref{fig-first-image}--\ref{fig-last-image}  as indicated in the 
    image.  The coordinates given are right ascension and declination
    relative to $\alpha_{2000}\!=\!00^{\rm h}26^{\rm m}35\fs70$,
    $\delta_{2000}\!=\!17\degr09\arcmin43\farcs06$. }     
  \label{fig-12k-field}
\end{figure*}

On 15 November 1999 we obtained a 3600~s V-band image using
the CFH12k camera (Cuillandre et al.\ \cite{CFH12k}) on CFHT (Figs.\
\ref{fig-12k-field} and
\ref{fig-first-image}-\ref{fig-last-image}). The six individual
exposures were bias and flat-field corrected in the standard way using
the \textsc{mscred} package under 
\textsc{Iraf}\footnote{IRAF is distributed by the
  National Optical Astronomy Observatories, which are operated by the
  Association of Universities for Research in Astronomy, Inc., under
  cooperative agreement with the National Science Foundation.}. 
The exposures were then registered onto the 
Digital Sky Survey\footnote{\texttt{http://archive.stsci.edu/dss/}}
%The Digitized Sky Surveys were
%produced at
%  the Space Telescope Science Institute under U.S. Government grant
%  NAG W-2166. The images of these surveys are based on photographic
%  data obtained using the Oschin Schmidt Telescope on Palomar Mountain
%  and the UK Schmidt Telescope. The plates were processed into the
%  present compressed digital form with the permission of these
%  institutions.
(DSS) image of the field and median combined. The final image, 
 a mosaic of all 12 chips with all the inter-chip gaps filled in,
has $\sim\!0\farcs7$ seeing (FWHM). The V-band photometric catalogue
obtained from this image contains $\sim\!3.7\times10^4$ objects on a
field of $\sim\!42\times28\,\mathrm{arcmin}^2$. 
 The internal errors on the V-band magnitudes are smaller than
  0.05 for $V\!<\!23.3$ (0.01 for $V\!<\!21.1$).
The limiting magnitude is $V\!\sim\!25$.  

In order to obtain colour information aperture magnitudes were
measured in $14\,\mathrm{pixel}$ ($2\farcs8$) diameter
apertures for the V- and I-band images. 
 This diameter is sufficiently
large compared to the seeing FWHM to enclose most of the light from
the object under consideration and small enough to avoid
contamination by neighbouring objects in crowded regions like the
cluster centre. The objects were then matched up using a polynomial
transformation of the I-band image coordinates onto the system of the
V-band image.  The centres of the two images coincide to within
40~arcsec, so that the overlapping region covers virtually
the whole UH8k field. Note that the gaps from the I band image show up
in the colour-magnitude catalogue as well. The resulting
colour-magnitude catalogue contains $\sim\!2.1\times10^4$
objects  with errors on the colours of smaller than 0.05 for
  $V\!<\!23.3$ (0.01 for $V\!<\!21.1$). Star-galaxy classification
over a sub-region of the colour-magnitude plane (as relevant for the
present paper) will be described in Sect.~\ref{ssec-completeness}.

A more detailed description of the photometric catalogue (including
star-galaxy separation over the whole colour-magnitude plane) is given
in Mayen et al.\ (\cite{mayen}) who use this catalogue to 
investigate the  depletion of background galaxies due to the
gravitational lens effect of Cl0024.

%%%%%%%%%%%%%%%%%%%%%%%% Spectroscopy %%%%%%%%%%%%%%%%%%%%%%
\subsection{Spectroscopy}
\label{ssec-spectroscopy}

Spectra were obtained using multi-slit spectroscopy during three
observing runs at CFHT and one at WHT. Table \ref{tab-obslog} shows
the observing log.  

Candidates for all the runs (except for run 1) were selected from 
a V-band mosaic obtained at the ESO New Technology Telescope (NTT) on
17/18 October 1993, with the primary selection criterion being 
$V_\mathrm{NTT}\!<\!23$. The seeing on this image was rather poor,
$\sim\!1\farcs7$ FWHM. In order to reduce contamination by stars,
the preparatory shallow R-band images were carefully examined during
all the CFHT runs; for runs 3 and 4, we took additional advantage of
the excellent seeing of the deep UH8k I-band image. 

All the CFHT observations were done with the Multi-Object Spectrograph
(MOS, Le F\`evre et al.\ \cite{mos}) with the O300 grism. The cameras
used during each run and the corresponding pixel scales and
dispersions are listed in Table \ref{tab-obslog}. Two  to five
exposures per mask were obtained, depending on the magnitudes of the
selected objects in each mask.    

Band-limiting filters, chosen such that prominent spectral features
(for instance the \ion{Ca}{i} H/K lines blueward of the 4000~\AA\
break) fall into 
the band at the redshift of interest, allow stacking of several rows of
spectra on one mask, thus increasing the number of spectra observable
in a given time. This strategy has been successfully employed by
e.~g.\ Yee et al.\ (\cite{CNOC}, \cite{CNOC2}). However, spectra
covering a larger range of wavelengths provide more secure redshift
determinations since more absorption and emission lines
can be taken into account. This is particularly important in the
presence of artifacts caused by insufficient removal of cosmic ray
hits. In the case of Cl0024+1654 at $z\!=\!0.395$, the
4000~\AA\ break roughly coincides with the strong sky emission line
[\ion{O}{i}]$\,\lambda$5577, so that insufficient subtraction of the sky
line might cause a problem in the redshift determination if only a
limited wavelength band were available. Also, covering a wide
wavelength range is essential in order to derive spectral types for the
galaxies and analyse in detail the spectral properties of the cluster
members. For these reasons we did not use
band-limiting filters and the usable wavelength range was typically
$4500$--$8500$~\AA, depending on signal-to-noise ratio and the quality
of the sky subtraction at the red end of the spectral range, where the
sky emission is dominated by molecular bands. 

The observations at WHT were made with the Low Dispersion Survey
Spectrograph (LDSS-2, Allington-Smith et al.\ \cite{ldss-2}), the
med/blue grism and the Loral LOR1 detector. The usable wavelength
range was $4000$--$7500$~\AA, somewhat bluer than for the MOS
observations. All the LDSS-2 masks were covered by two exposures each. 

We used $1\arcsec$ wide slits throughout, resulting in a resolution
of $\sim\! 13$~\AA, except for run 1 where the slit width
used was $1\farcs5$ with a correspondingly worse resolution of $\sim\! 
20$~\AA.

%%%%%%%%%% observing log table %%%%%%%%%%%%%%%%%%%%%%%%%%%%
\begin{table*}
  \caption{
    Log of the spectroscopic observing runs.
    }
  \begin{tabular}{clllcccc@{\hspace{-0.1mm}}ccl}
    \hline\noalign{\smallskip}
    \# & Date & Instrument & CCD & Grism & $N_{\rm masks}$ & Exp.\
    time & \multicolumn{2}{c}{Pixel} & Disp. & Std. star \\ 
    & & & & & & (ksec) & ($\mu\mathrm{m}$) & (arcsec) & (\AA/pix) & \\
    \noalign{\smallskip}
    \hline\noalign{\smallskip}
    
    1 & 24--27/08/92 & CFHT/MOS & SAIC1 & O300     & 2 & 4.5 -- 7.5  &
    18 & 0.377 & 4.31 & Feige 110 \\
    2 & 24--27/08/95 & CFHT/MOS & Loral-3 & O300     & 6 & 6.6 -- 15.6
    & 15 & 0.314 & 3.69 & Wolf 1346 \\
    3 & 12--15/09/96 & WHT/LDSS-2  & LOR1 & med/blue & 9 & 5.4 -- 7.2
    & 15 & 0.357 & 3.30 & HZ4\\
    4 & 11--13/11/96 & CFHT/MOS & STIS-2 & O300-1   & 3 & 6.6 -- 8.1 &
    21 & 0.440 & 5.03 & Hiltner 600 \\
    \hline\noalign{\smallskip}
    \noalign{\smallskip}

%% for 1 and 2 could add BD 25 (39 41),but not enough space. 

  \end{tabular}
  \label{tab-obslog}
\end{table*}

%%%%%%%%%%%%%%%% data reduction %%%%%%%%%%%%%%%%%%%%%%%%%%%%
\section{Data reduction and analysis}
\label{sec-reduction}

\subsection{Reduction of spectroscopic data}
\label{ssec-reduction}

The spectra were reduced using the semiautomatic package
\textsc{multired} (Le F\`evre et al.\ \cite{multired}) which in turn
uses standard \textsc{Iraf} tasks and treats each slit separately. The
spectral images were de-biased and flat-field corrected in the
standard way. A low-order (mostly linear) polynomial fit to the sky
emission in the spatial direction was then subtracted for each
individual exposure. During the fit most of the cosmic ray hits in
the sky region of the spectral images were taken care of by a
sigma-rejection algorithm (pixels with deviations of more than
$\pm2\sigma$ from the fit were rejected before refitting);
nevertheless insufficiently rejected cosmics occasionally survived the
fit to produce fake absorption features in the object spectrum.    

Geometric distortions cause the true spatial/dispersion directions to
deviate from the row/column directions of the CCD chip, especially at
the edges of the field. Since sky fitting was done row-wise
(column-wise for the WHT data), sky emission lines tend to be
imperfectly subtracted. Isolated emission lines (in particular
[\ion{O}{i}]$\lambda 5577$, which roughly coincides with the 4000~\AA\ 
break for galaxies at the redshift of Cl0024+1654) could simply be
masked for the subsequent analysis of the spectrum, but the usable
wavelength range in the red was effectively limited by the molecular
band emission from the sky.  

The individual exposures were then averaged into the final
two-dimensional spectrum. In those cases where more than two exposures
per mask were available, the highest pixel value was rejected, thus
accounting for cosmic ray hits. The small number of exposures per mask
made cosmic ray rejection difficult and imperfect. A cosmic ray hit on
the object spectrum resulted in a fake emission feature, a hit in the
sky area in a fake absorption feature. However, as argued in Section
\ref{ssec-redshift}, cosmics do not in general influence the redshift
determination. Finally, variance-weighted one-dimensional spectra were
extracted from the combined spectral images, wavelength calibrated
using lamp spectra (He/Ar at CFHT, Cu/Ar at WHT) and approximately
flux-calibrated with long-slit spectra from spectro-photometric
standard stars (listed in Table \ref{tab-obslog}).  

The wavelength calibration spectra were extracted using
a straight trace (following the column/row direction of the CCD),
unlike the corresponding object spectra, where the 
trace followed the flexure introduced by geometric camera
distortions. We verified that this causes only negligible errors by
re-calibrating several spectra with strong flexure with calibration
spectra extracted using the same trace as for the object spectra. 

We finally performed a check on the wavelength calibration by 
extracting three spectra per mask without sky subtraction and
measuring the position of prominent sky lines using the calibration
from the lamps. For the WHT masks this revealed systematic shifts,
which are due to the fact that the lamp spectra were not
taken immediately before or after the science exposures. The
positioning of the masks in the mask holder was therefore not
identical during lamp and science exposures. These systematic shifts
translate to $\sim\!10^{-3}$ in terms of redshift and a correction for
each mask was applied to  all the redshifts determined from this
mask. No such systematic effect was found for the CFHT masks.  

\begin{figure}
  \resizebox{\hsize}{!}{\includegraphics{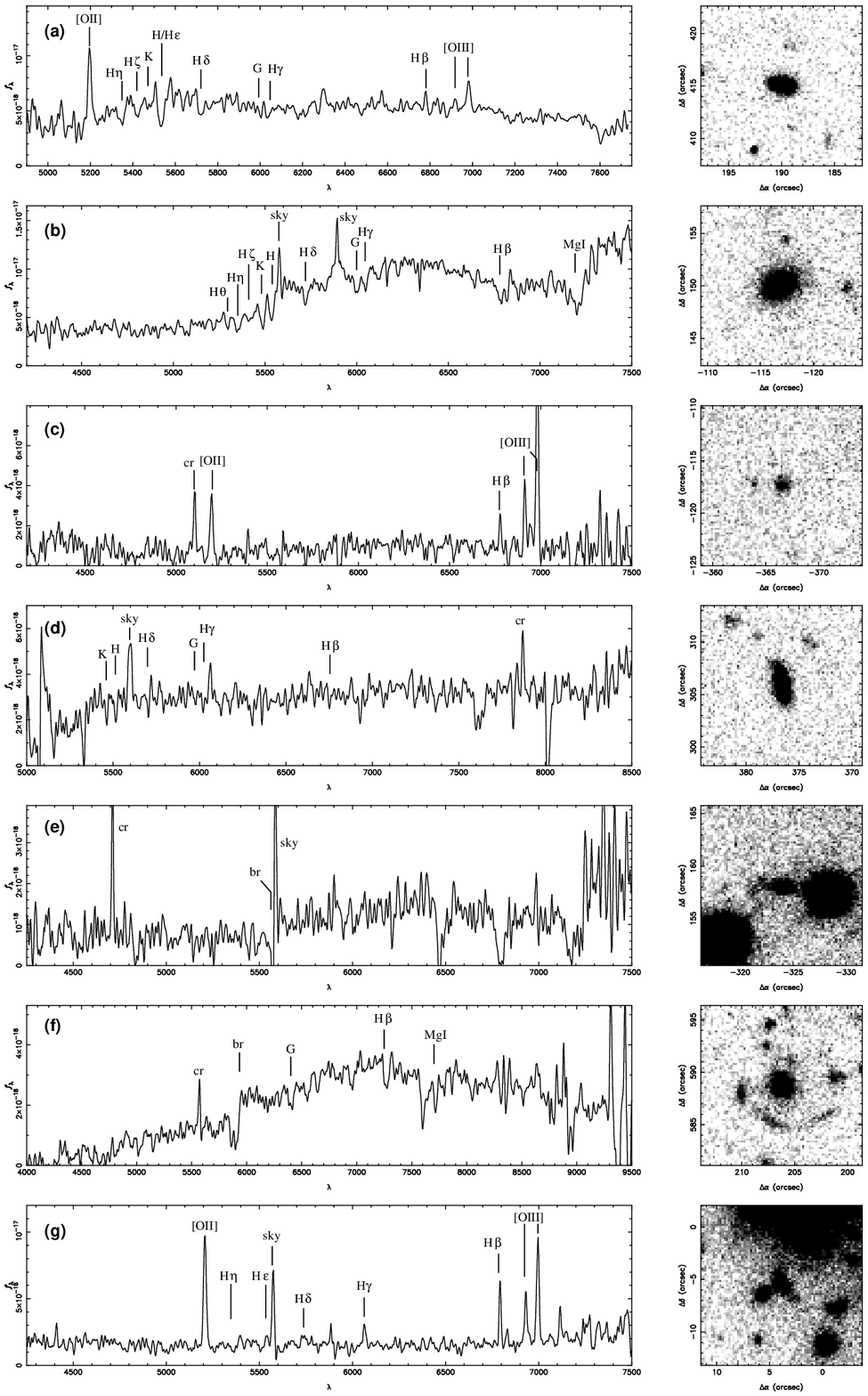}}
%  \resizebox{0.9\hsize}{!}{\includegraphics{figures/expl-spectra.ps}}
  \caption{Example spectra and corresponding
    $15\!\times15\,\mathrm{arcsec}^2$ sections from the CFH12k V-band
    image. The catalogue entries for these objects are listed at the
    top of Table \ref{tab-catalog}. Spectra (a) and (d) are from
    observing run 2 (CFHT), spectra (b), 
    (c), (e) and (g) from run 3 (WHT), and spectrum (f) is from run 4
    (CFHT). Spectra (a), (b) and (c) are examples of spectra with
    ``secure/A'' redshifts based on a large number of emission and/or
    absorption lines. Spectrum (d) is an example of a ``probable/B''
    redshift and spectrum (e) of an ``uncertain/D'' redshift, based on 
    the 4000~\AA\ break only. Spectrum (f) is for the central galaxy
    of the northern group at $z\!\sim\!0.495$, discussed in Sect.\
    \ref{sec-discussion}. Spectrum (g), finally, is for a very 
    blue cluster member near the centre of Cl0024+1654, showing almost
    the complete Balmer series in emission. Cosmic ray hits are marked
    ``cr'', sky emission lines by ``sky''. Note that the wavelength
    ranges are different in each panel.}
  \label{fig-expl-spectra}
\end{figure}

Seven example spectra and the V-band images for the corresponding
objects are shown in Fig.\ \ref{fig-expl-spectra}.

%%%%%%%%%%%%%%%%%% redshift identification %%%%%%%%%%%%%%5
\subsection{Redshift determination}
\label{ssec-redshift}

Since a large fraction of the spectra in our sample have rather low
signal-to-noise ratio, we decided to identify redshifts by eye, which is
better at finding real absorption and emission lines amongst noise
than automatic redshift identification techniques. Also, fake emission
and absorption features due to cosmics could be identified this way
by simply checking the original two-dimensional spectral images. 
All redshifts were identified by one of us (OC for runs 2-4, GS for
run 1) and checked by at least one other member of the team. The
redshifts (runs 2-4 only) given in Table \ref{tab-catalog} carry a
flag which indicates a (somewhat subjective)  level of confidence that
the redshift identification is correct. ``Secure/A'' redshifts were 
determined from spectra where emission lines and/or several absorption
lines are clearly seen, ``uncertain/D'' redshifts are based on a
tentative identification of a single line or possibly several weak
absorption lines, and ``probable/B'' and ``possible/C'' indicate
intermediate levels of confidence.  

In order to obtain a more objective estimate of the \emph{measurement} (as
opposed to \emph{identification}) error of our redshifts, we used the
cross-correlation technique implemented in the task \textsc{xcsao}
in the \textsc{Iraf} package \textsc{rvsao} (Kurtz \& Mink
\cite{rvsao}).  In all cases, the by-eye redshift identification was
fed to \textsc{xcsao} as the initial redshift estimate and the allowed
redshift range was restricted to about $\pm 5$\% around this
redshift. In this sense, \textsc{xcsao} was forced to recover our
redshift identification and we only used \textsc{xcsao}'s error
estimate. 

For large values of the correlation parameter, $R\! > \! 3$ (Kurtz \&
Mink \cite{rvsao}), we find that \textsc{xcsao} reproduces our input
redshifts very well. The mean deviation between by-eye redshift and
\textsc{xcsao} redshift is $\overline{\left(z_\mathrm{eye} \! - \!
  z_\mathrm{xcsao}\right)}\! =\! -0.00018$ with a scatter of 0.00037,
corresponding to 80~km~s$^{-1}$ at the redshift of
Cl0024. However, only 107 of our spectra achieve $R\! > \! 3$ and the
redshift distribution of these is not representative of the redshift
distribution of the total sample, low-redshift objects being more
reliably identified than objects at the cluster redshift or beyond. 
\textsc{xcsao} assigns an individual error estimate to each
redshift, based on the width of the peak in the correlation
function. For $R\! > \! 3$, the median of the distribution of this error
is at $\sim \! 0.0003$ (43~km~s$^{-1}$), with the bulk of error
estimates at $<\! 0.0005$. 

\begin{table}[tbp]
  \caption{Comparison of multiply observed objects. The first three
    lines give numbers determined from objects observed twice during
    the same runs, the last three lines compare different runs. All
    the redshift differences are given in units of $10^{-4}$. 
    The numbers of the observing runs are those given in
    Table \ref{tab-obslog}.} 
 
 \begin{tabular}{r@{ -- }ll@{\hspace{0.3cm}}rcr@{$\pm$}ll@{\hspace{0.3cm}}rc}
    \hline\noalign{\smallskip}
    \multicolumn{2}{c}{} & & \multicolumn{4}{c}{all qualities} & & \multicolumn{2}{c}{\hspace*{-0.5cm}``secure/A'' only} \\
    \multicolumn{2}{l}{Runs} & & $N$ &
    $\overline{\left|z_1-z_2\right|}$ &
    \multicolumn{2}{r}{$\overline{\left(z_1-z_2\right)}$} & & $N$ & 
    $\overline{\left|z_1-z_2\right|}$ \\
    \noalign{\smallskip}\hline\noalign{\smallskip}
    2 & 2 & & 19 & 8.2 & \multicolumn{2}{c}{---} & & 19 & 8.2 \\
    3 & 3 & &  6 & 4.8 & \multicolumn{2}{c}{---}& & 4 & 5.8 \\
    4 & 4 & &  9 & 6.6 & \multicolumn{2}{c}{---}& & 7 & 5.4 \\
    \noalign{\smallskip}\hline\noalign{\smallskip}
    2 &   3 & & 28 & 14.0 &  $6$ & $19$ & & 26 & 10.6 \\
    2 & D99 & & 24 & 12.8 &  $2$ & $19$ & & 22 &  9.6 \\
    3 & D99 & & 28 & 10.0 & $-6$ & $16$ & & 25 & 13.4 \\
    \noalign{\smallskip}\hline
  \end{tabular}
\label{tab-redundants}
\end{table}

However, these error estimates are only useful for our ``best''
spectra. In order to get a more reliable error estimate for all the
spectra we can compare redshifts determined from multiple observations
either during the same run, when the same object appears on different
masks, or during different observing runs. The results of this
intercomparison are shown in Table \ref{tab-redundants} for those
pairs of runs with useful number $N$ of pairs: The mean absolute
differences between the redshifts are of order $1\!\times\!10^{-3}$,
with no marked difference between the
values determined for all quality codes or ``secure/A'' redshifts
only. We note that the slits at the ``southern'' ends of the WHT masks
were consistently of rather poor quality, in the sense that the slit
edges become quite rugged. Redshifts from these slits are therefore
less accurate than those from the central and ``northern'' parts of
the WHT masks or from the observing runs at CFHT. Taking this fact
into account we estimate that most of our redshifts are accurate to
about $\sim\!1\!\times\!10^{-3}$. The majority of our spectra were
obtained during observing runs 2 and 3; comparing the redshifts of
objects observed during both these runs we find that the systematic
shift $\overline{\left(z_1-z_2\right)}$ between the observations is 
consistent with zero.  

\begin{figure}
  \resizebox{\hsize}{!}{\includegraphics{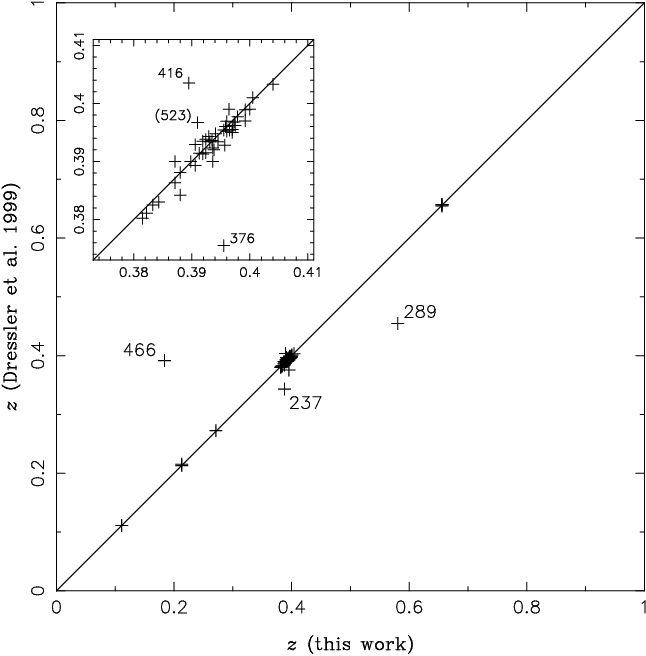}}
%  \resizebox{0.9\hsize}{!}{\includegraphics{figures/z_comparison.ps}}
  \caption{Comparison of redshift measurements for objects observed
    both by us and by Dressler et al.\ (\cite{dressler2}). The large
    panel shows the full samples, the inset is a blow-up of the
    cluster region. Five clear misidentifications and one $3\sigma$
    drop-out are labeled with their numbers in the spectroscopic
    catalogue. The sizes of the symbols in the inset correspond to
    half our redshift error estimate of $\pm1\times10^{-3}$.  
}
  \label{fig-DG_comparison}
\end{figure}

 Finally we compare in Fig.\ \ref{fig-DG_comparison} the redshifts
  for those 54 objects that were observed both by us and by Dressler
  et al.\ (\cite{dressler2}). For the majority of these
  objects the redshift measurements agree very well, however, there are
  five clear misidentifications, with redshift deviation of more than
  0.01; these cases are discussed in detail in Sect.\
  \ref{ssec-catalogue}. Object 523 deviates by more than $3\sigma$ if
  the five misidentifications are dropped from the sample, so we exclude
  this object as well. The remaining 48 common objects have a mean
  redshift deviation of $\overline{z_{\rm D99} \!-\! z_{\rm our}}\!=\!
  -0.0003$ with a root mean square (RMS) scatter of $\sigma\!=\!0.0015$. The
  rms scatter is consistent with the estimate for our redshift error
  estimate of $10^{-3}$, assuming that the redshifts given by Dressler et
  al.\ (\cite{dressler2}) have similar accuracy. Again, there is no
  evidence for a systematic shift.

%%%%%%%%%%%%%%%%%% Spectroscopic measures %%%%%%%%%%%%%%
\subsection{Spectroscopic measures}
\label{ssec-spectroscopic}

We measured equivalent widths for [\ion{O}{ii}]$\,\lambda 3727$,
[\ion{O}{iii}]$\,\lambda\lambda 4959,5007$, H$\alpha$ (where within the
wavelength range), H$\beta$ and H$\delta$, as well as the strength of
the 4000\AA\ break. [\ion{O}{iii}], H$\alpha$ and H$\beta$ were measured
semi-automatically, i.\ e.\ the continuum level was placed by visual
inspection and the line integration limits were fixed at the values
given in Table \ref{tab-ew-ranges}. The integration ranges for
[\ion{O}{iii}] and H$\beta$ are the same as those used by Dressler \&
Shectman (\cite{ewranges}); the range for H$\alpha$ is taken
from Couch et al.\ (\cite{halpha}) and does not include the
neighbouring [\ion{N}{ii}] line. 

[\ion{O}{ii}] and H$\delta$ are important indicators of ongoing
(Kennicutt \cite{kennicutt}) and recently terminated (Abraham et al.\
\cite{abraham}) star formation within galaxies and in particular provide
essential information on the interaction of newly accreted galaxies
with the cluster environment. We therefore adopted a more accurate way
to determine the equivalent widths and in particular
to estimate a level of significance for the strengths of these
lines.  We define our equivalent widths so that they are
\emph{positive} for emission lines:
\begin{equation}
  \label{eq:ew-definition}
  W_\lambda 
  = \sum_{i=1}^{N_{\rm int}} \frac{f_i - \overline{f_{\rm
        c}}}{\overline{f_{\rm c}}}\,\Delta\lambda  
  = \sum_{i=1}^{N_{\rm int}} \frac{f_i}{\overline{f_{\rm c}}} -
  N_{\rm int}\,\Delta\lambda \quad\mbox{,}
\end{equation}
where $f_i$ is the flux in pixel $i$, $N_{\rm int}$ the number of pixels
in the integration range, and $\Delta\lambda$ is the wavelength
dispersion in \AA/pixel. The continuum level 
$\overline{f_{\rm c}}$ was estimated as  
the mean value within two wavelength intervals on either side of the
line; pixels with values more than $3\sigma$ away from the mean level
were iteratively rejected in order to avoid cosmic ray hits in
the continuum region. The automatic measurement of the continuum level
makes it possible to estimate an error on the corresponding equivalent
widths, thus allowing an  assessment of  the detection significance of the
line. For this purpose we model the noise as Poisson-distributed photon
noise. This allows us to relate the variances on $f_i$ and
$\overline{f_{\rm c}}$ to the single-pixel signal-to-noise ratio at
flux level $\overline{f_{\rm c}}$, $S/N = \overline{f_{\rm
    c}}/\sigma_{\rm c}$ using the 
usual Poisson scaling $\sigma^2\propto f$ and 
$\sigma\left(\overline{f_{\rm c}}\right)^2 = \sigma_{\rm c}^2/N_{\rm
  c}$ when averaged over $N_{\rm c}$ pixels. Adding the errors due to
$f_i$ and $\overline{f_{\rm c}}$ in quadrature we thus obtain:
\begin{eqnarray}
  \lefteqn{\sigma_{W_\lambda}^2 =} \nonumber\\
  & & \left(\frac{S}{N}\right)^{-2} 
  \left[ \left(W_\lambda + N_{\rm int}\,\Delta\lambda\right)\Delta\lambda +
    \frac{\left(W_\lambda + N_{\rm
          int}\,\Delta\lambda\right)^2}{N_{\rm c}} 
  \right].
\label{eq-OII-error}
\end{eqnarray}
Here the first term is due to the line integration and the second to
the determination of the continuum level.    

In the case of
[\ion{O}{ii}], the signal-to-noise ratio $S/N$ was determined between
3560~\AA\ and 
3680~\AA, a region which is largely free from absorption lines. For
H$\delta$, the signal-to-noise ratio was measured in the range 4050~\AA\ to
4250~\AA, where the line itself was excised between 4085~\AA\ and
4115~\AA. The latter $S/N$ was measured for every spectrum as a global
indicator of the quality of the spectrum. The integration ranges
correspond to those used by Abraham et al.\ (\cite{abraham}). For
H$\delta$ we used their ``narrow'' range.  

The strength of the 4000~\AA\ break is given as the ratio of the total
flux in the range $4050$~\AA$ < \! \lambda \! < \! 4250$~\AA\ to the
total flux within $3750$~\AA$ < \! \lambda \! < \! 3950$~\AA. These
ranges include all the absorption lines. 

All the wavelength ranges are in the rest-frame of the object. For
spectra for which no redshift could be determined we only estimated
the global $S/N$ assuming the cluster redshift $z\!=\!0.395$. 

\begin{table}
  \caption{
    Wavelength ranges for equivalent width measurements. All the
    wavelengths are given in \AA\ (``i/a'' --- interactive placement of
    continuum level).  
    }
  \begin{tabular}{rcccc}
    \hline\noalign{\smallskip}
    & $\lambda_\mathrm{cent}$ & line & blue cont. & red cont. \\
    \noalign{\smallskip}
    \hline\noalign{\smallskip}
    [\ion{O}{ii}]  & 3727 & 3713--3741 & 3653--3713 & 3741--3801 \\[0em]
    [\ion{O}{iii}] & 5007 & 4997--5017 & i/a & i/a \\
    H$\alpha$      & 6563 & 6556--6570 & i/a & i/a \\
    H$\beta$       & 4861 & 4851--4871 & i/a & i/a \\
    H$\delta$      & 4103 & 4088--4116 & 4030--4082 & 4122--4170 \\
    break          & 4000 &  --        & 3750--3950 & 4050--4250 \\
    S/N            & ---  & 4050--4250 &  --        &  -- \\
    \noalign{\smallskip}
    \hline\noalign{\smallskip}
  \end{tabular}
  \label{tab-ew-ranges}
\end{table}

\stepcounter{figure}
\newcounter{savefig}
\setcounter{savefig}{\value{figure}}
\setcounter{figure}{0}
\renewcommand{\thefigure}{\mbox{\arabic{savefig}\alph{figure}}}
\begin{figure}
%%% for draft: no images
% \resizebox{\hsize}{!}{\includegraphics[draft]{figures/cl0024_1.ps}}
%%% for astro-oh: binned images
%%  \resizebox{\hsize}{!}{\includegraphics{figures/smcl0024_1.ps}}
%%% for publication: full images
  \resizebox{\hsize}{!}{\includegraphics{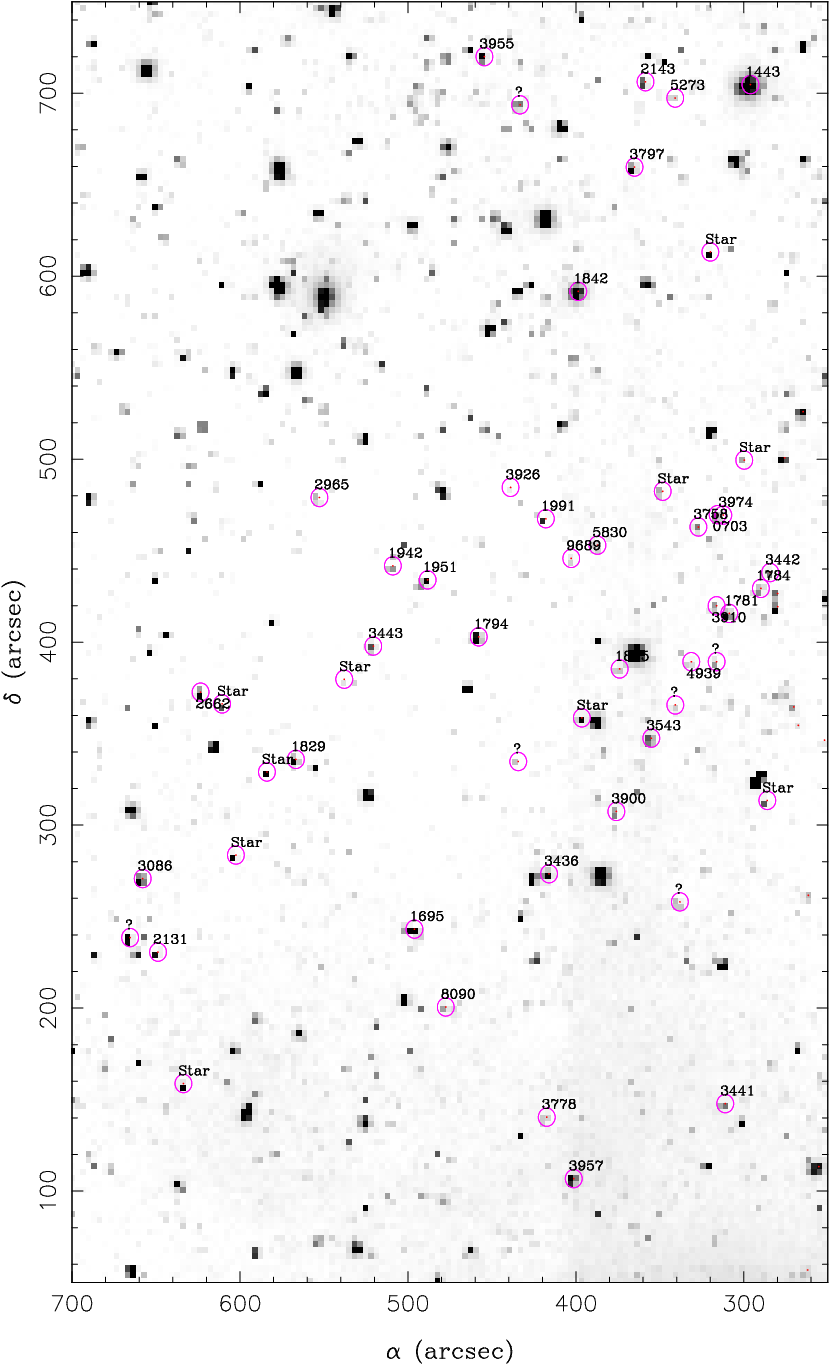}}
  \caption{Subsections of the CFH12k V-band image with redshifts for
    members of the spectroscopic sample. The positions of the
    subsections on the whole field are marked in Fig.\
    \ref{fig-12k-field}. For redshifts $z\!<\!1$ only the decimals are
    given; for galaxies marked `?' spectra are available but no
    redshift could be determined. The (electronic) catalogue (cf.\
    Table \ref{tab-catalog}) is sorted by relative right ascension, so
    the entry for any object can be easily identified by noting its
    coordinates.} 
  \label{fig-first-image}
\end{figure}

\begin{figure}
%%% for draft: no images
%  \resizebox{\hsize}{!}{\includegraphics[draft]{figures/cl0024_2.ps}}
%%% for astro-oh: binned images
%%  \resizebox{\hsize}{!}{\includegraphics{figures/smcl0024_2.ps}}
%%% for publication: full images
  \resizebox{\hsize}{!}{\includegraphics{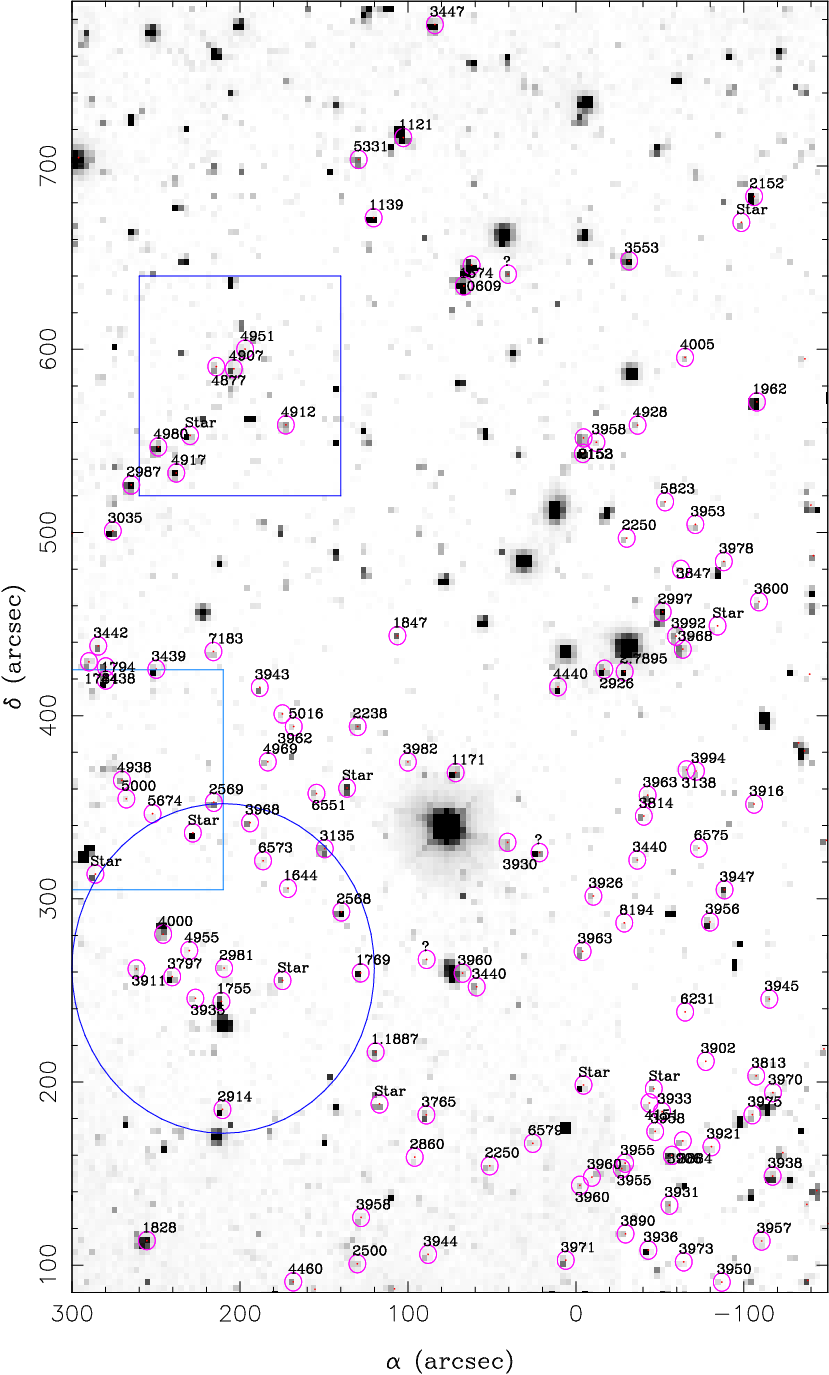}}
  \caption{Continued. The squares mark the positions of Figs.\
    \ref{fig-group0.49}, the circle the area around the weak shear
    signal detected by Bonnet, Mellier \& Fort (\cite{BMF}), discussed
    in Sect.\ \ref{sec-discussion}.}
  \label{fig-image2}
\end{figure}

\begin{figure}
%%% for draft: no images
%  \resizebox{\hsize}{!}{\includegraphics[draft]{figures/cl0024_3.ps}}
%%% for astro-oh: binned images
%%  \resizebox{\hsize}{!}{\includegraphics{figures/smcl0024_3.ps}}
%%% for publication: full images
  \resizebox{\hsize}{!}{\includegraphics{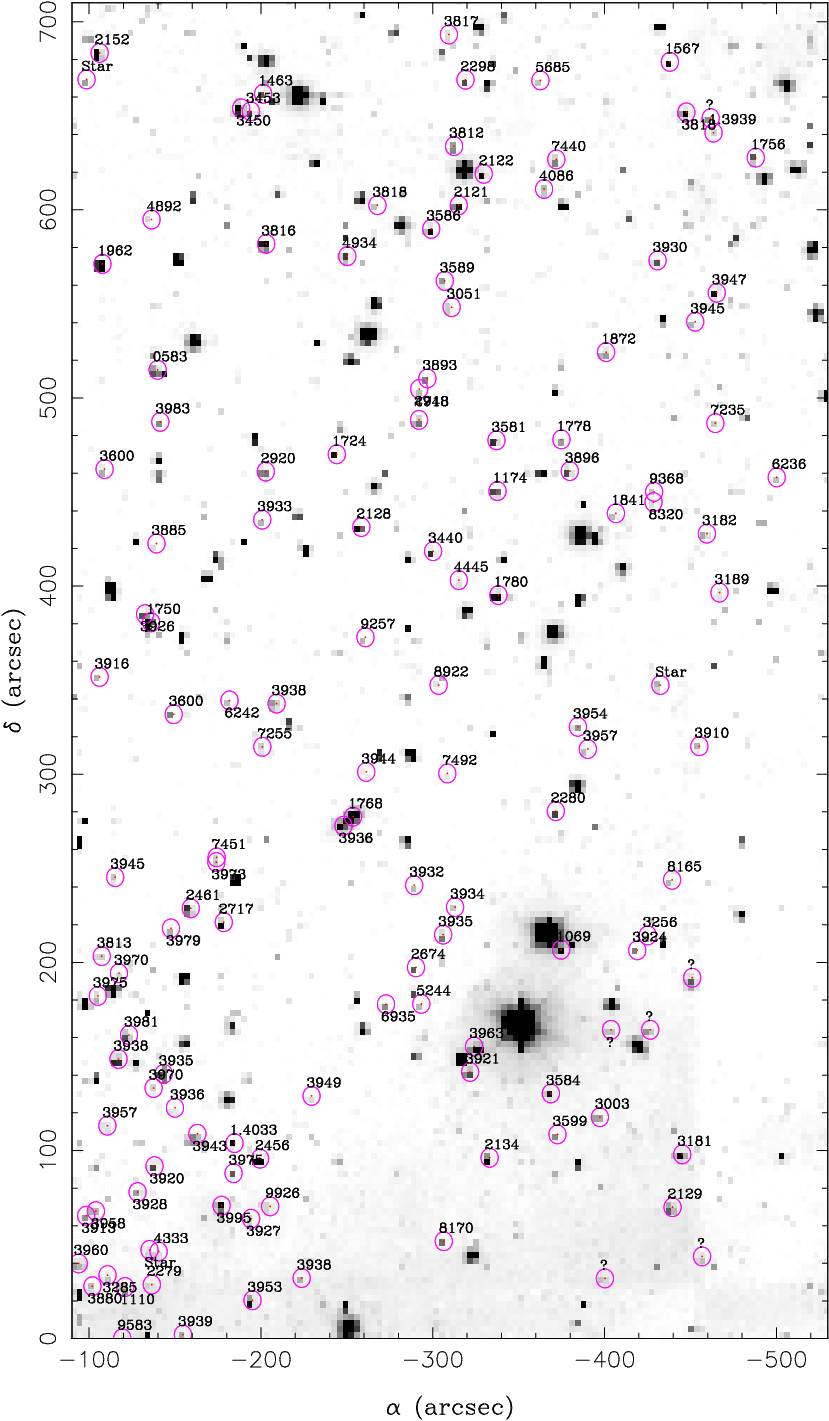}}
  \caption{Continued}
  \label{fig-image3}
\end{figure}

\begin{figure}
%%% for draft: no images
%  \resizebox{\hsize}{!}{\includegraphics[angle=270,draft]{figures/cl0024_4.ps}}
%%% for astro-oh: binned images
%%  \resizebox{\hsize}{!}{\includegraphics[angle=270]{figures/smcl0024_4.ps}}
%%% for publication: full images
  \resizebox{\hsize}{!}{\includegraphics[angle=270]{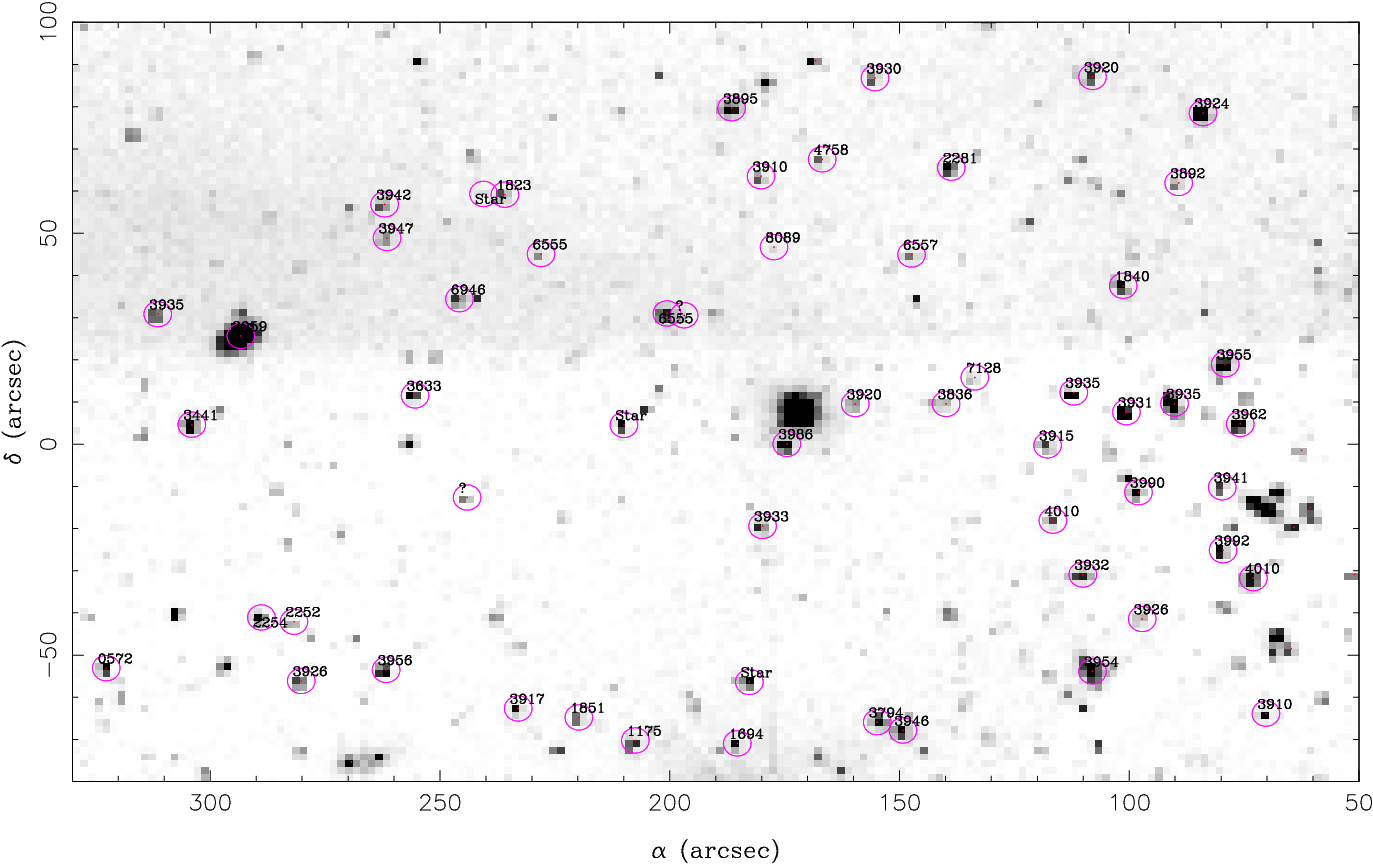}}
  \caption{Continued}
  \label{fig-image4}
\end{figure}

\begin{figure}
%%% for draft: no images
%  \resizebox{\hsize}{!}{\includegraphics[draft]{figures/cl0024_5.ps}}
%%% for astro-oh: binned images
%  \resizebox{\hsize}{!}{\includegraphics{figures/smcl0024_5.ps}}
%%% for publication: full images
  \resizebox{\hsize}{!}{\includegraphics{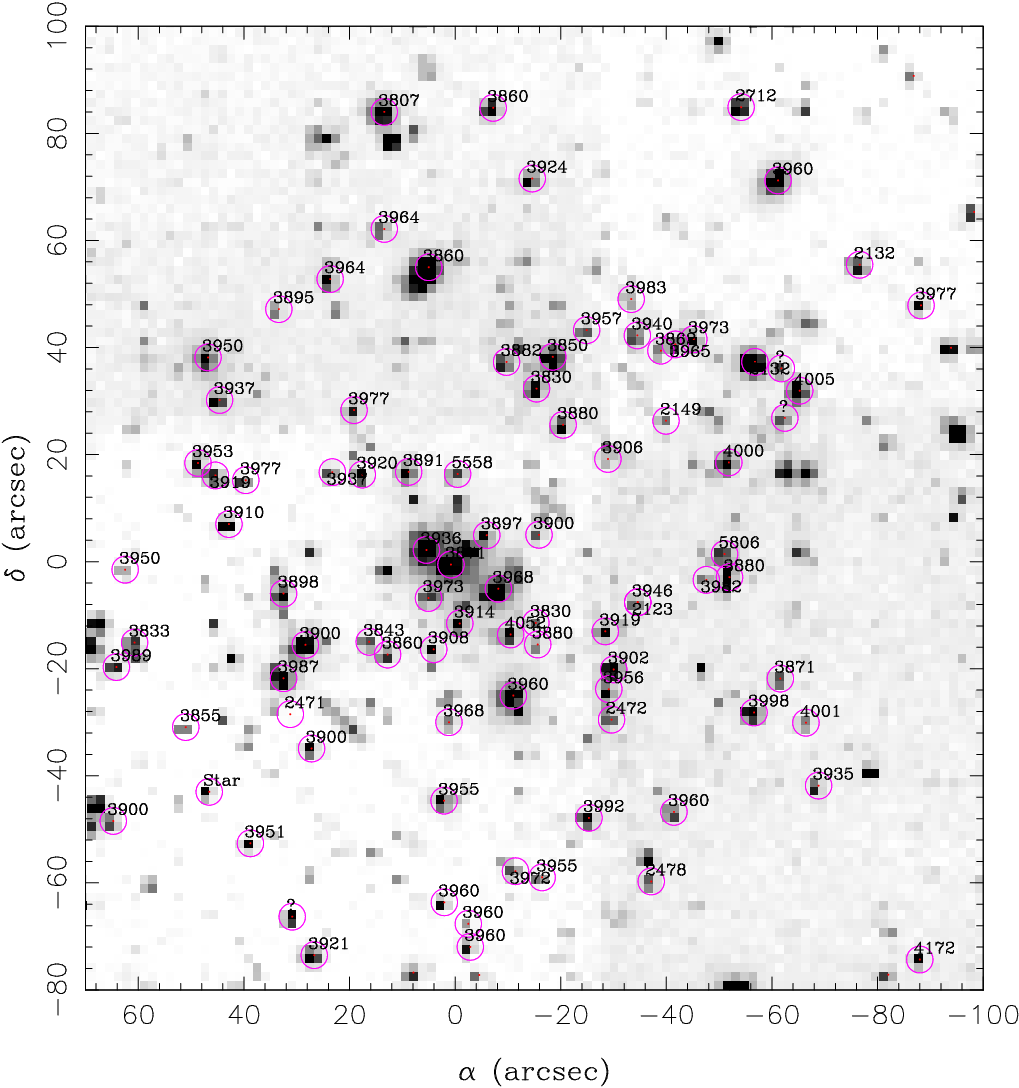}}
  \caption{Continued}
  \label{fig-image5}
\end{figure}

\begin{figure}
%%% for draft: no images
%  \resizebox{\hsize}{!}{\includegraphics[angle=270,draft]{figures/cl0024_7.ps}}
%%% for astro-oh: binned images
%  \resizebox{\hsize}{!}{\includegraphics[angle=270]{figures/smcl0024_7.ps}}
%%% for publication: full images
  \resizebox{\hsize}{!}{\includegraphics[angle=270]{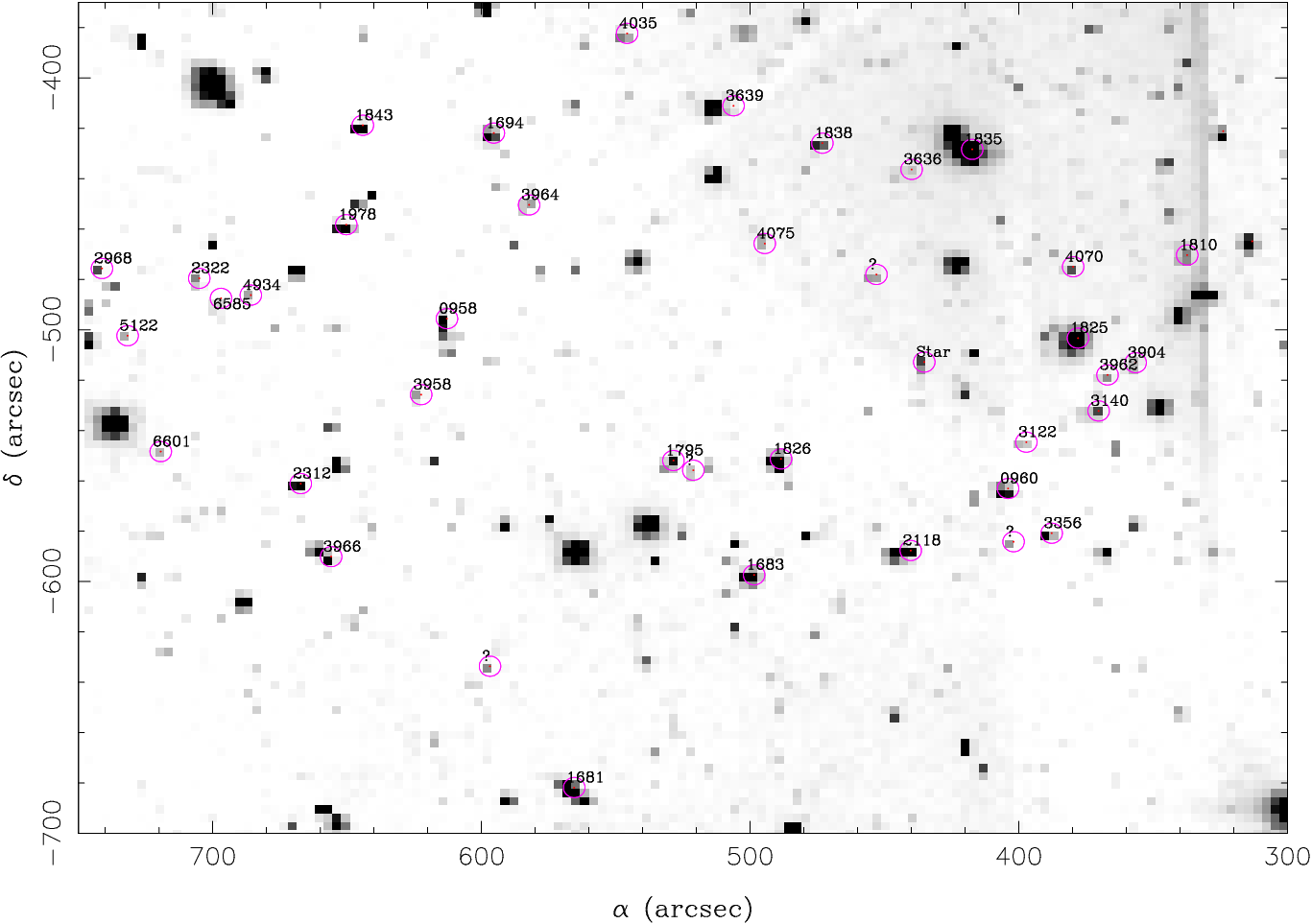}}
  \caption{Continued}
  \label{fig-image7}
\end{figure}

\begin{figure}
%%% for draft: no images
%  \resizebox{\hsize}{!}{\includegraphics[draft]{figures/cl0024_8.ps}}
%%% for astro-oh: binned images
%  \resizebox{\hsize}{!}{\includegraphics{figures/smcl0024_8.ps}}
%%% for publication: full images
  \resizebox{\hsize}{!}{\includegraphics{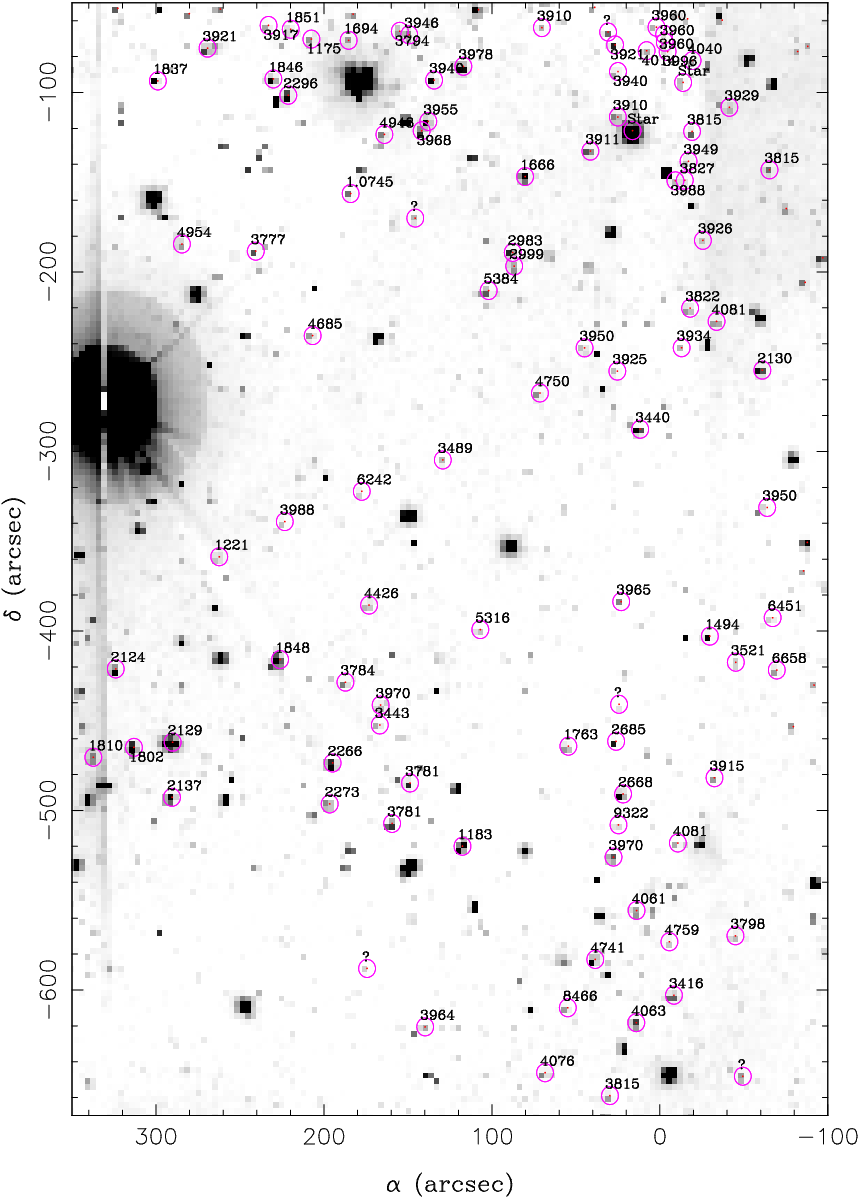}}
  \caption{Continued}
  \label{fig-image8}
\end{figure}

\begin{figure}
%%% for draft: no images
%  \resizebox{\hsize}{!}{\includegraphics[draft]{figures/cl0024_9.ps}}
%%% for astro-oh: binned images
%  \resizebox{\hsize}{!}{\includegraphics{figures/smcl0024_9.ps}}
%%% for publication: full images
  \resizebox{\hsize}{!}{\includegraphics{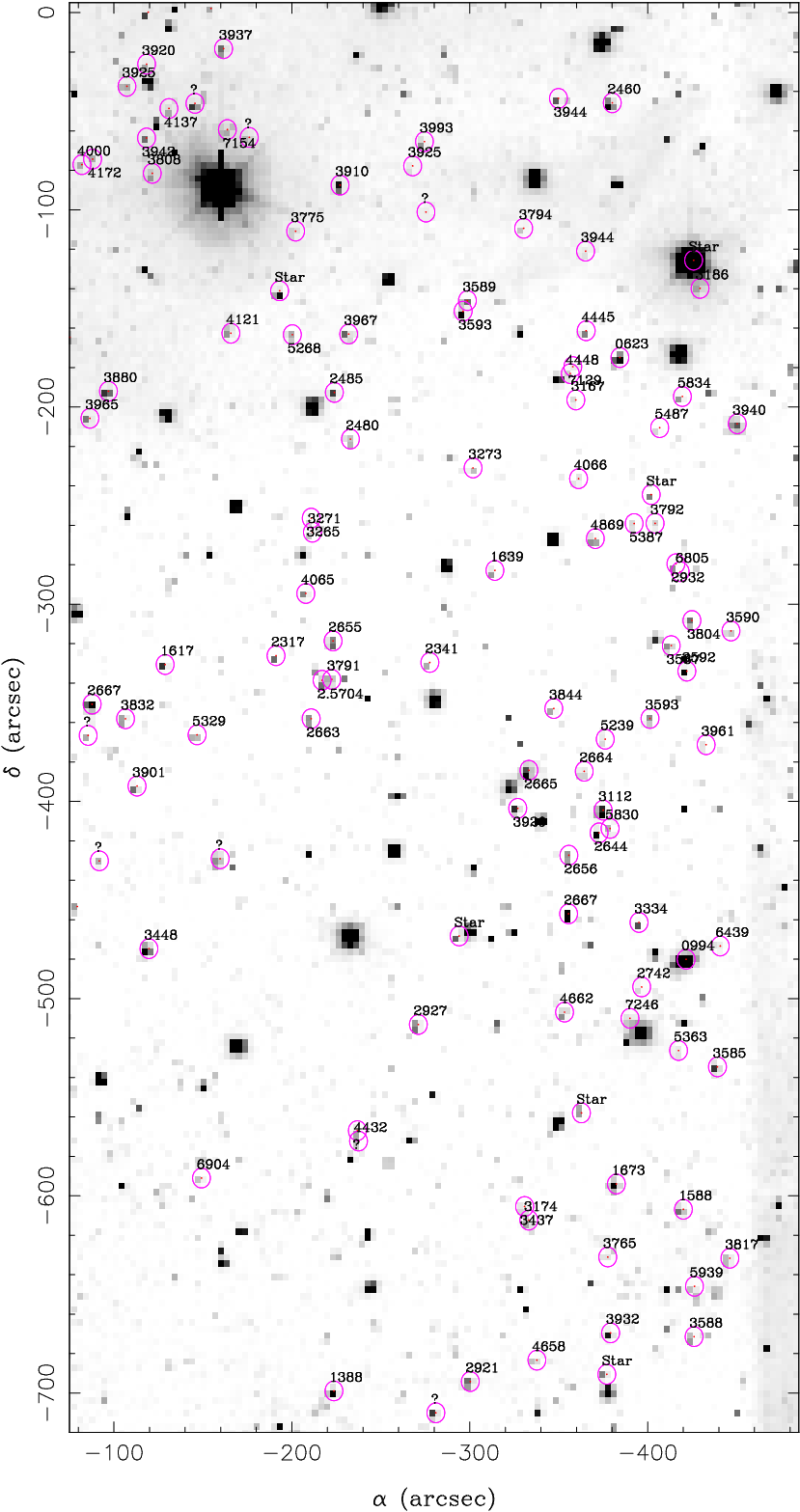}}
  \caption{Continued}
  \label{fig-last-image}
\end{figure}

\setcounter{figure}{\value{savefig}}
\renewcommand{\thefigure}{\arabic{figure}}

%%%%%%%%%%%%%%%%%%%% Catalogue %%%%%%%%%%%%%%%%%%%%
\subsection{The catalogue}
\label{ssec-catalogue}

An excerpt from the final catalogue is shown for reference in Table
\ref{tab-catalog}, the full catalogue is available in electronic
form 
 at the Centre de Donn\'ees Stellaire
  (CDS)\footnote{\texttt{http://cdsweb.u-strasbg.fr/cats/J.A+A.htx}}. All 
entries from the catalogue are marked with their redshifts in Figs.\
\ref{fig-first-image}-\ref{fig-last-image}.

%%%%%%%%%%%%%%% the catalogue %%%%%%%%%%%%%%%%%%%%%%%%%%%%%%%%
\begin{sidewaystable*}
  \caption{The catalogue. This table presents a selection from the
    sample, the full catalogue is available in electronic form at
    CDS. The first seven entries correspond to the example spectra
    shown in Fig.\ \ref{fig-expl-spectra}. The next seven/six entries 
    belong to the northern/southern groups at $z\!\sim\!0.495$,
    discussed in Sect.\ \ref{sec-discussion}. Finally we list
    several objects in the vicinity of the dark structure detected by
    Bonnet, Mellier \& Fort (\cite{BMF}). See the text for a
    description of the table contents.}
  \include{bigtable}  
  \label{tab-catalog}
\end{sidewaystable*}

In detail the contents of Table \ref{tab-catalog} are as follows:
        
\noindent \textbf{Column 1:} Object number. The catalogue is sorted by
relative right ascension (Column 2). 

\noindent \textbf{Column 2/3:} Right ascension and declination relative to
$\alpha\!=\!00^{\rm h}26^{\rm m}35\fs70$, $\delta\!=\!17\degr09\arcmin43\farcs06$ (J2000), given in arcsec.        

\noindent \textbf{Column 4:} Redshift.

\noindent \textbf{Column 5:} Redshift reliability code: A =
``secure'', B = ``probable'', C = ``possible'', D = ``uncertain'', S =
``star''. For objects taken from Dressler et al.\ (\cite{dressler2})
and not observed by us, we give their quality code (ranging from 1 to
4). 

\noindent \textbf{Column 6:} $V$ magnitude (SExtractor
\texttt{MAG\_BEST}), from CFH12k image.

\noindent \textbf{Column 7:} $V\!-\!I$ colour, measured in 14 pixel
($2\farcs8$) diameter apertures from the CFH12k ($V$) and UH8k ($I$)
images. Occasionally $I$ magnitudes (and hence $V\!-\!I$ colour) are
not available due to the object falling on a gap between two chips of
the UH8k (see Sect.\ \ref{ssec-imaging}).    

\noindent \textbf{Column 8:} [\ion{O}{ii}]$\lambda3727$ equivalent
width (in \AA). The error was estimated using Eq.\ \ref{eq-OII-error}.
We use the convention that equivalent widths are positive for emission
and negative for absorption lines. For objects taken from Dressler et
al.\ (\cite{dressler2}) and not observed by us, we remeasured the
equivalent widths ourselves from their spectra, so as to provide
homogeneous equivalent width measurements. 

\noindent \textbf{Column 9:} [\ion{O}{iii}]$\lambda5007$ equivalent
width (in \AA). 

\noindent \textbf{Column 10:} H$\alpha$ equivalent width (in \AA).

\noindent \textbf{Column 11:} H$\beta$ equivalent width (in \AA).
XS
\noindent \textbf{Column 12:} H$\delta$ equivalent width (in \AA). The
error was estimated using Eq.\ \ref{eq-OII-error}.

\noindent \textbf{Column 13:} Strength of the 4000~\AA\ break.

\noindent \textbf{Column 14:} Signal-to-noise ratio measured between
4050 and 4250~\AA. 

\noindent \textbf{Column 15:} Number(s) of the observing run(s) during
which the object was observed (cf.\ Table \ref{tab-obslog}). Objects
observed by Dressler et al.\ (\cite{dressler2}) are marked by ``D''
and the number from their catalogue. 

\noindent \textbf{Column 16:} Lines that could be identified in the
spectrum/spectra.

We provide new spectra for 618 objects, of which 581 have
redshifts. The global success rate is therefore 94\%. Not all
redshifts are equally secure though: We qualify 435 of our 
redshifts (70\%) as ``secure'' (quality code A), 28 (5\%) as
``uncertain'' (code D) and 86 (14\%) of intermediate quality. 34
objects (5\%) turned out to be stars. The fraction of ``secure''
redshifts is 83\% for foreground galaxies at $z\!<\!0.37$, 82\% for
galaxies around the cluster redshift ($0.37\!<z\!<0.41$), and drops to
53\% for galaxies at higher redshift. 
 
For completeness, we include in the catalogue the redshifts provided
by Dressler et al.\ (\cite{dressler2}). The corresponding
entry number from their catalogue (prefixed by `D') is listed in
column 15 of our catalogue. The catalogue as published by Dressler et
al.\ (\cite{dressler2}) contains 130 entries, of which 107 are cluster
members. However, we noticed several errors in their catalogue,
reducing the number of distinct objects to 125\footnote{ 
  Dressler et al.\ (\cite{dressler2}) failed to make several
  identifications within their sample: thus D19 and D66 seem to be the
  same object, as are D21 and D65, D113 and D120, D122 and D125, D22
  and D73.  
}. 54 of these objects were observed by us as
well, usually with concordant redshifts, see the discussion in Sec.\
\ref{ssec-redshift} and Fig.\ \ref{fig-DG_comparison}. There were
however five problematic 
cases: Dressler et al.\ give the redshift for object 376 (D7) 
as 0.3755, whereas our spectrum indicates a secure 0.3955. We assume
that this is a typographical error in the Dressler et al.\ list and
adopt our value. Our spectrum for object 416 (D33) indicates
$z\!=\!0.3895$ as opposed to their 0.4035. Since Dressler et al.\ only
give this a quality code `3', we adopt our value. Object 289 (D109) is
similarly uncertain.  Object 466 should be identical to Dressler et
al.'s  object D37, however the spectra are completely different giving
secure redshifts of 0.3916 and 0.1840 respectively. We have to assume 
therefore that D37 has wrong coordinates. A similar problem occurs for
D130 (our 237).   

Adding the 69 objects which were observed by Dressler et
al.\ alone increases the size of the catalogue to 687 objects, with 650
identified redshifts.

%%%%%%%%%%%%%%%%%%% Completeness %%%%%%%%%%%%%%%%%%%%
\subsection{Completeness}
\label{ssec-completeness}

\begin{figure}
  \resizebox{\hsize}{!}{\includegraphics{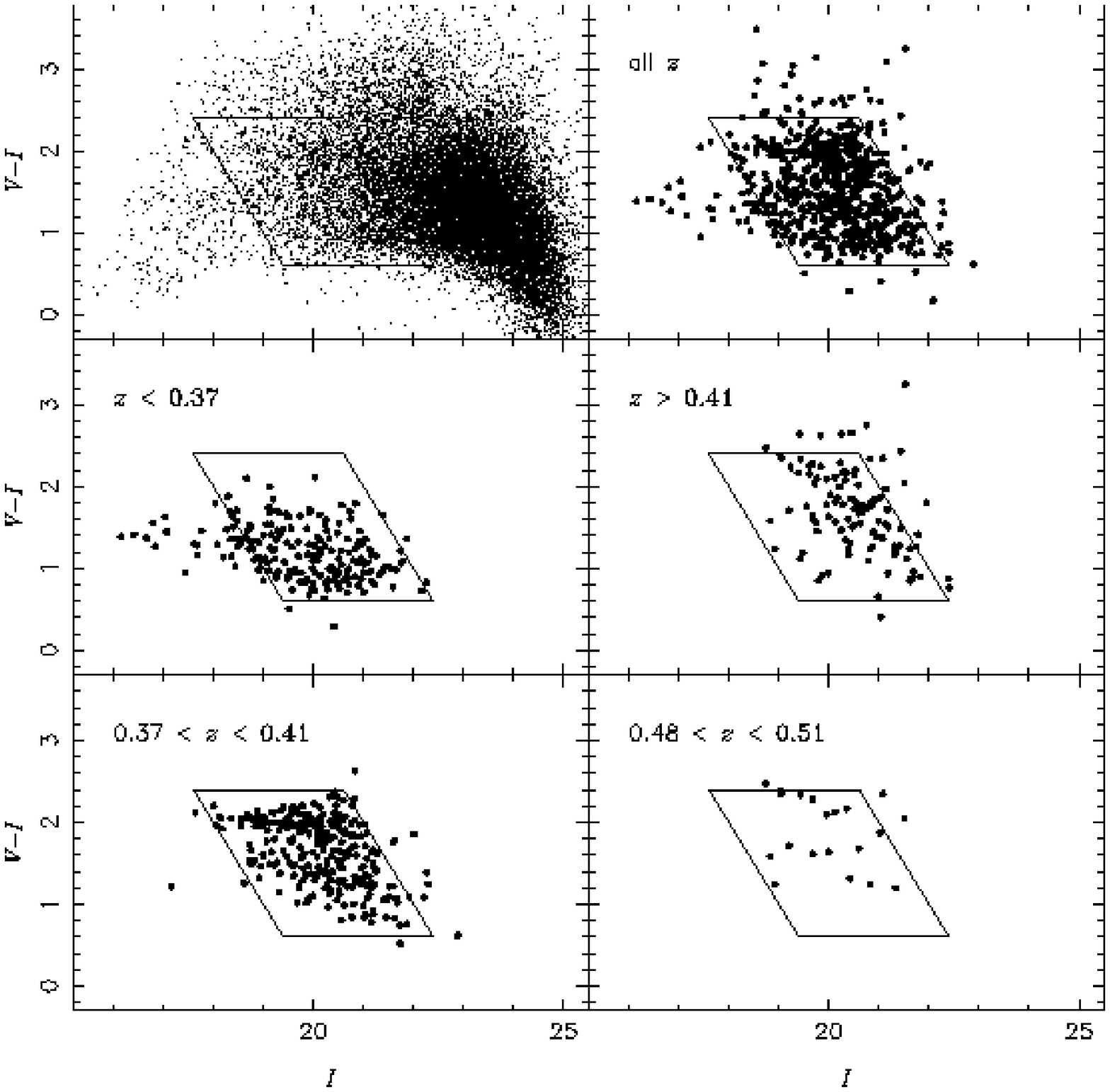}}
  \caption{$V$-$I$ colour-magnitude diagrams. The top-left diagram
    shows the full photometric catalogue, the top-right diagram the
    full spectroscopic catalogue. The next three diagrams split
    the spectroscopic catalogue according to redshift, showing
    foreground and background galaxies as well as galaxies around the
    cluster redshift $z\!\sim\!0.39$. Note the clearly visible cluster
    sequence at $V\!-\!I \!\simeq\! 2$ in the latter diagram. The
    bottom-right diagram shows the members of the newly discovered
    group of galaxies at $z\!\sim\!0.495$ (see Section
    \ref{sec-discussion}). The parallelogram marks the
    subsample used in the completeness map (Fig.\
    \ref{fig-completeness-map}), $20 \!<\! V \!<\! 23$, $0.6 \!<\!
    V\!-\!I \!<\! 2.4$.   
    }
  \label{fig-colmag}
\end{figure}

Fig.\ \ref{fig-colmag} presents colour-magnitude diagrams (CMDs) for
the photometric (for the overlap region of the CFH12k and UH8k images)
and the spectroscopic catalogues. For the latter catalogue we also
show separate CMDs for foreground galaxies at $z\!<\!0.37$, background
galaxies at $z\!>\!0.41$, galaxies at the cluster redshift
($0.37 \!<\! z \!<\! 0.41$) and for a newly identified group of
galaxies at $z\!=\!0.495$ (see Sect.\ \ref{sec-discussion}).  

\begin{figure}
  \resizebox{\hsize}{!}{\includegraphics{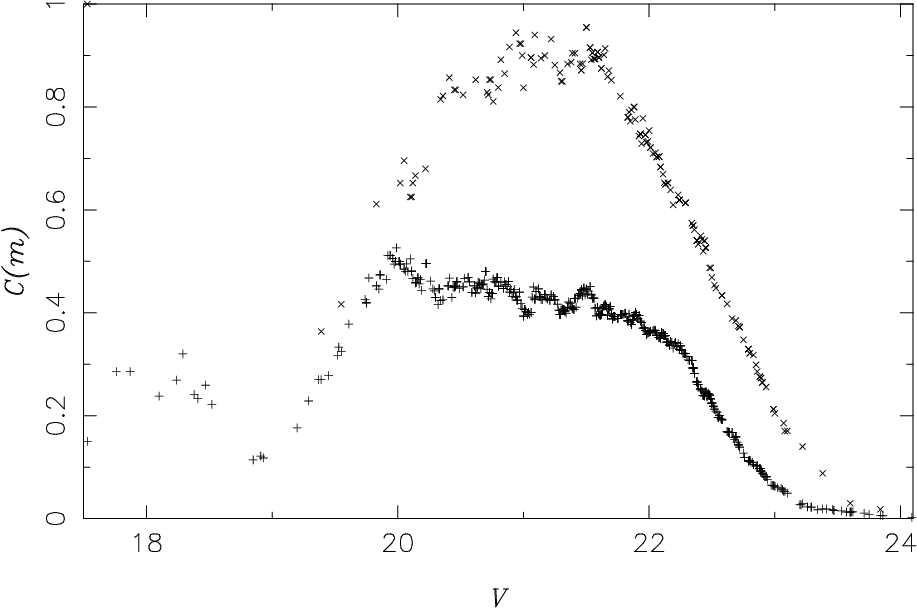}}
  \caption{Completeness of the spectroscopic survey in $V$
    magnitude. For each galaxy from the catalogue this is given as the
    ratio of the numbers of galaxies in the spectroscopic 
    and photometric catalogues in a given bin width centered on the
    magnitude of the galaxy. Pluses mark galaxies taken from
    the whole survey area (as outlined in Fig.\
    \ref{fig-completeness-map}), crosses galaxies within $3\arcmin$ of
    the cluster centre. In the former case, a bin width of $0.5$~mag
    was used, in the latter a bin width of 1~mag. 
    }
  \label{fig-mag_completeness}
\end{figure}

\begin{figure}
  \resizebox{\hsize}{!}{\includegraphics{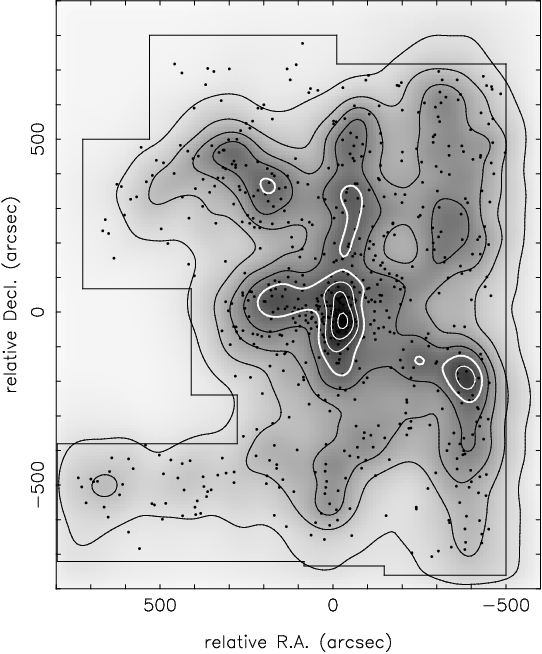}}
  \caption{Map of the completeness variation of the spectroscopic
    catalogue as gray-scale with overlaid contours. The completeness
    at any point is determined in a circular top-hat encompassing the
    10 nearest neighbours in the spectroscopic survey; the map is
    smoothed with a Gaussian of width $30\arcsec$. Contour lines
    are spaced in 10\% steps. The 50\% contour is marked by a bold
    line, contours at less than 50\% are drawn in black, higher
    contours in white.  
    }
  \label{fig-completeness-map}
\end{figure}

The primary criterion for the candidate selection for the
spectroscopic survey was $V\!<\!23$. %$V_\mathrm{NTT}\!<\!23$. 
The completeness in $V$ magnitude of the final catalogue is shown in
Fig.\ \ref{fig-mag_completeness}, where for each galaxy from the 
spectroscopic catalogue we count the number of objects in a bin of 
$\pm0.25\,\mathrm{mag}$ around the galaxy magnitude in both the
spectroscopic and photometric catalogues (the latter restricted to the
survey area as outlined in Fig.\ \ref{fig-completeness-map}) and
define their ratio as the total completeness at the given
magnitude. Between $V\!\simeq\!20$ and $V\!\simeq\!22$ the 
completeness is roughly constant at $\sim\!45\,$\% and drops rapidly
for fainter magnitudes. Restricting the same analysis to the central
area within $3\arcmin$ of the cluster centre (using larger bins of
$\pm0.5\,\mathrm{mag}$) shows that in this region the completeness
exceeds 80\% between $V\!\simeq\!20$ and $V\!\simeq\!22$. 

A visual impression of the variation of the completeness
across the survey area is given in Fig.\ \ref{fig-completeness-map}
where we use an adaptive top-hat (its radius at any given point of a
grid is the geometric mean of
the distances to the 10th and 11th nearest objects with spectroscopy)
to compute the ratio of the numbers of objects with spectroscopy
(fixed to 10) and the number of objects in the photometric
catalogue. In order to have samples with well-defined (though
  \textit{a posteriori}) photometric selection limits
  we restrict the spectroscopic and photometric catalogues for this
  purpose to subsamples with $20\!<\!V\!<\!23$ and $0.6 \!<\! V\!-\!I
  \!<\! 2.4$. As Fig.~\ref{fig-colmag} shows, this region of the
  colour-magnitude plane encloses virtually all the cluster members in
  the spectroscopic catalogue. Within our restricted sample the
  separation between stars and   galaxies within the photometric
  catalogue is fairly straightforward using SExtractor's
  \texttt{CLASS\_STAR} parameter. Within a total of 2722 
  objects we find 312 stars (with an error of about $\pm10$ due to
  ambiguous cases) or 11.5\%. Since useful shape information was not
  available for the preparation of all the spectroscopic observing
  runs (about 5\% of the objects in the spectroscopic catalogue are
  stars), the whole photometric sample is used in 
  Fig.~\ref{fig-completeness-map}. The completeness values are
  therefore actually \textit{underestimates} if interpreted as
  completeness for spectroscopic coverage of \textit{galaxies} alone.
The cluster centre is sampled at $>\!70$\% completeness in  a region
of about $80\arcsec\!\times\!150\arcsec$.

%%%%%%%%%% colour magnitude diagrams %%%%%%%%%%%%%%%%%%%%

%%%%%%%%%%%%%%%%%%%% magnitude completeness %%%%%%%%%%%%%%%%%%%%

%%%%%%%%%%%%%%%%%%%% completeness map %%%%%%%%%%%%%%%%%%%%

%%%%%%%%%%%%%%%%%%%%%%%%%%%%%%%%%%%%%%%%%%%%%%%%%%%%%%%%%%
%%%%%%%%%%%%%%%%%%%%%%%% Global Properties %%%%%%%%%%%%%%%
%%%%%%%%%%%%%%%%%%%%%%%%%%%%%%%%%%%%%%%%%%%%%%%%%%%%%%%%%%
\section{Discussion}
\label{sec-discussion}

\begin{figure}
  \resizebox{\hsize}{!}{\includegraphics{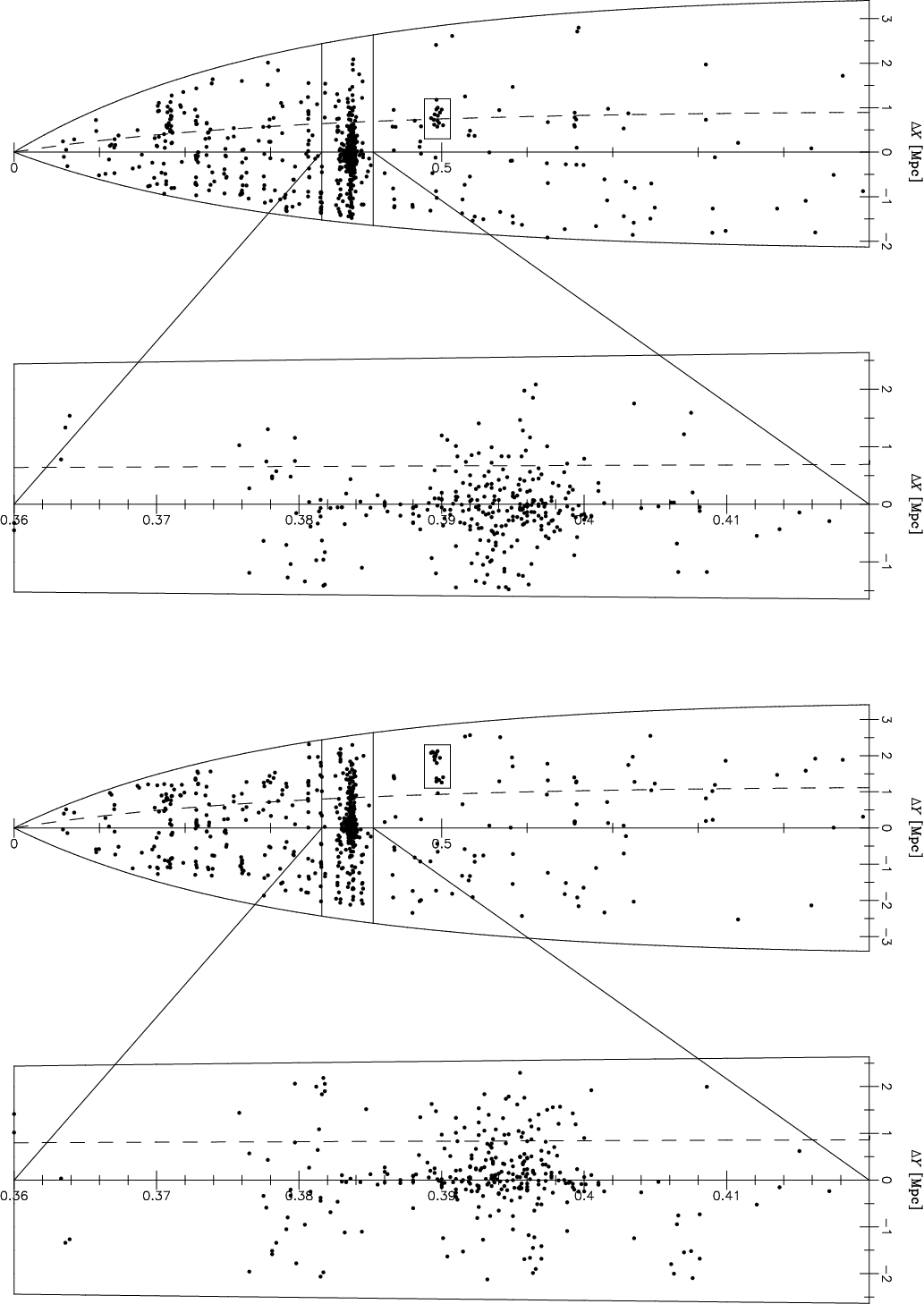}}
  \caption{Three-dimensional distribution of the objects in our
    redshift catalogue. In the two upper panels the objects are
    projected onto the right ascension axis, in the lower two onto the
    declination axis. The upper panel of each pair shows the
    large-scale distribution from $z\!=\!0$ to $z\!=\!1$, the lower
    panel an expanded view of the environment of the cluster Cl0024
    itself. The dashed line marks the direction towards the potential
    perturbation detected by Bonnet et al.\ (\cite{BMF}). Two groups
    at $z\!\sim\!0.495$ are marked by rectangles. The
    conversion from angular position on the sky to proper 
    transverse distance was done assuming an Einstein-de Sitter
    Universe with
    $H_0\!=\!100\,\mathrm{km}\,\mathrm{s}^{-1}\,\mathrm{Mpc}^{-1}$.}  
  \label{fig-wedge}
\end{figure}

The three-dimensional distribution of the galaxies in the redshift
sample is shown in Fig.\ \ref{fig-wedge} in the form of wedge
diagrams, where the angular position of each object on the sky has
been converted to proper distance from the line of sight, appropriate
for the given redshift in a $\Omega_{\rm M}\!=\!1$,
$\Omega_\Lambda\!=\!0$ world model. The cluster Cl0024+1654 shows up
clearly as a sheet at $z\!\simeq\!0.4$. The expanded views show that
Cl0024+1654 is not a simple isolated cluster but that there is a foreground
``clump'' at $z\!=\!0.38$, superimposed onto the main cluster and
connected to the latter via a narrow bridge. In a companion paper
(Czoske et al.\ \cite{paper2}) we discuss an interpretation of this
structure as being due to the foreground cluster having passed through
the main cluster. 

We find a pair of compact groups of galaxies at $z\!=\!0.495$ about
$10\arcmin$ to the north-east of the centre of Cl0024+1654 (see Fig.\
\ref{fig-wedge} at $x\!\simeq\!1$~Mpc, $y\!\simeq\!2$~Mpc). The northern
group includes 8 galaxies, centered at 
$574\arcsec$ north and $203\arcsec$ east of the cluster centre, the
southern group includes 6 galaxies centered at $369\arcsec$ north and
$250\arcsec$ (median positions); the projected distance between the
groups is thus $\sim\!740\,h^{-1}\,\mathrm{kpc}$. The mean redshifts
are $\overline{z_{\rm N}}\!=\!0.4921$ and $\overline{z_{\rm
S}}\!=\!0.4970$, the formal velocity dispersions $\sigma_{\rm
N}\!=\!657\,{\rm km}\,{\rm s}^{-1}$ and $\sigma_{\rm S}\!=\!647\,{\rm
km}\,{\rm s}^{-1}$. Student's t-test rejects the hypothesis  that
the two groups have the same mean redshift at 99\% confidence, so we
assume that we are really seeing two separate groups. The velocity 
dispersions are presumably enhanced by tidal interaction between the
groups.  

\begin{figure}
%%% fuer draft
%  \resizebox{\hsize}{!}{\includegraphics[draft]{figures/groupeN.ps}}
%%% fuer publication
%  \resizebox{\hsize}{!}{\includegraphics{groupeN.ps}}
%%% fuer referee
%  \resizebox{0.8\hsize}{!}{\includegraphics{figures/groupeN.ps}}
  \vspace{0.5cm}
%%% fuer draft
% \resizebox{\hsize}{!}{\includegraphics[draft]{figures/groupeS.ps}}
%%% fuer publication
% \resizebox{\hsize}{!}{\includegraphics{groupeS.ps}}
%%% fuer referee
% \resizebox{0.8\hsize}{!}{\includegraphics{figures/groupeS.ps}}
  \caption{``True'' colour images of the apparent centres of the
    northern (top) and southern (bottom) groups of galaxies at
    $z\!\simeq\!0.49$. The images were created from the V and I band
    images, the green channel is an average of these two
    images. Supposed member galaxies of the group are conspicuous by
    their yellow colour, corresponding to $V\!-\!I\!\sim\!2.4$. Note
    the blue arc-like structure around the galaxy at $z\!=\!0.4907$.}  
  \label{fig-group0.49}
\end{figure}

Fig.\ \ref{fig-group0.49} shows colour images of the northern and
southern groups created from the I- and V-band CCD images. The galaxy at 
$z\!=\!0.4907$ is surrounded by three objects of similar, blue colour.
It is tempting to interpret this group as multiple images of the same
background object. In this case, using the curvature radius
($\sim\!5\arcsec$) as an estimate for the Einstein radius and $z_{\rm
  s}\!=\!1$ as a rough guess for the redshift of the background
source, we obtain $5.8\times10^{12}\,h^{-1}\,M_{\sun}$ for the mass
within this radius.  

Another overdensity in Fig.~\ref{fig-wedge} occurs at
$z\!\sim\!0.18$. These galaxies are however distributed fairly uniformly
across the field with no obvious spatial concentration and are
therefore just part of the general large-scale structure in the Universe. 

In the first detection of a coherent shear field around a cluster of
galaxies, Bonnet et al.\ (\cite{BMF}) found a signal to the north-east
of the centre of Cl0024+1654, indicating a concentration of mass at a
point where no overdensity of galaxies is apparent in the
two-dimensional images. The direction to this dark ``clump'' is
indicated by a circle in Fig.~\ref{fig-image2} and by the dashed line
in Fig.\ \ref{fig-wedge}. There is no significant over-density
along this line which could explain the spatial tightness of the
signal observed by Bonnet et al.

%%%%%%%%%%%%%%%%%%%%%%%%%%%%%%%%%%%%%%%%%%%%%%%%%%%%%%%%%%
%%%%%%%%%%%%%%%%%%%%%%%% Conclusions %%%%%%%%%%%%%%%%%%%%%
%%%%%%%%%%%%%%%%%%%%%%%%%%%%%%%%%%%%%%%%%%%%%%%%%%%%%%%%%%
\section{Conclusions}
\label{sec-conclusions}

In this paper we present a catalogue of spectroscopic data for 687
objects in a field of $21\!\times\!25\,\mathrm{arcmin}^2$ around the
centre of the cluster of galaxies Cl0024+1654 at
$z\!\sim\!0.395$, 295 of which lie in the vicinity of the cluster
itself, in the range $0.37\!<\!z\!<\!0.41$. The completeness of the
sample exceeds 80\% around the cluster centre for $V\!<\!22$, dropping
to $\sim\!70$\% for $V\!<!23$. The mean completeness over the whole
field is about 45\% down to $V\!=\!22$. This is therefore one
of the largest spectroscopic surveys available for a cluster at
$z\!\ga\!0.2$ and the largest at $z\!\sim\!0.4$. Apart from redshifts
the catalogue lists photometric data and equivalent widths for five
lines as well as the strength of the 4000~\AA\ break, important for
determining present and recent star forming activity. The
catalogue represents a unique database for investigations into the
structure of a medium-redshift cluster of galaxies and its environment. 

We report the discovery of a binary group of galaxies at
$z\!\simeq\!0.495$, which includes a gravitational arc candidate.  A
further overdensity in the redshift distribution at $z\!\sim\!0.18$
can at present not be attributed to a collapsed structure, although we
cannot exclude the possibility that a centre for such a structure
exists outside the field covered by our survey. None of the structures
seen in the redshift distribution seems to be able to explain the
coherent shear signal at $\sim\!6\arcmin$ to the north-east of the
projected centre of Cl0024+1654, detected by Bonnet et al.\
(\cite{BMF}). 

The main result of this project is that Cl0024+1654 is not a simple
isolated cluster, as has hitherto been assumed in interpreting
kinematical, lensing and X-ray data. Instead there is a second, less
massive cluster projected onto the centre of the main cluster. The
separation of Cl0024+1654 into two components has strong consequences
for the interpretation of observational data and should in particular
help to resolve the well-known discrepancy between masses determined
using different methods for this cluster (Soucail et al.\
\cite{soucail}). The structure of Cl0024+1654, a scenario for possible
interaction between the main and foreground clusters and the effects
of this interaction on the properties of the cluster galaxy
populations will be discussed in more detail in the second paper of
this series (Czoske et al.\ \cite{paper2}). The photometric data
obtained in conjunction with this project will be used by Mayen et
al.\ (\cite{mayen}) to investigate the depletion of background objects
due to the gravitational lensing effect of Cl0024+1654.

%%%%%%%%%%%%%%%%%%%% Acknowledgements %%%%%%%%%%%%%%%%%%
\begin{acknowledgements}
  OC thanks the European Commission for generous support under grant
  number ER-BFM-BI-CT97-2471. JPK thanks CNRS for support and
  acknowledges a  European Large Scale Facility grant for the  
  WHT observation. This work was supported by the TMR Network 
  ``Gravitational Lensing: New Constraints on Cosmology and the
  Distribution of Dark
  Matter''\footnote{\texttt{http://www.ast.cam.ac.uk/IoA/lensnet}} 
  of the European Commission under contract No.\ ER-BFM-RX-CT97-0172. 
  TJB wishes to thank Observatoire Midi-Pyr\'en\'ees for hospitality.  
\end{acknowledgements}

%%%%%%%%%%%%%%%%% References %%%%%%%%%%%%%%%%%%%%%%

\end{document}

%% file: bigtable.tex
%#\begin{tabular}[h]{rrrrlrrrrrrrrrl}
\begin{tabular}[h]{rrrclcccccccccrl}
\hline\noalign{\smallskip}
num & R.A. & Dec & $z$ & Q & $V$ & $V-I$  & [\ion{O}{ii}] & [\ion{O}{iii}] & H$\alpha$ & H$\beta$ & H$\delta$ & 4000~\AA & $S/N$ & Runs & Lines \\
\noalign{\smallskip}
\hline\hline
\noalign{\smallskip}
531 & 189.8 & 414.9 & 0.3943 & A & 21.70 & 1.18 & $23\pm$3 & 6 & - & 4 & $-4\pm$1 & 1.1 & 11.7 & 2 & H/K, H$\zeta$, H$\eta$, H$\theta$, [\ion{O}{ii}]\dots\\
221 & -117.1 & 150.3 & 0.3938 & A & 20.46 & 1.73 & - & - & - & - & $-3\pm$1 & 1.8 & 16.9 & 3, D97 & br, H/K, H$\zeta$, H$\eta$, H$\theta$\dots\\
 80 & -366.5 & -117.5 & 0.3944 & A & 23.46 & 0.62 & $77\pm$15 & 109 & - & 23 & - & 1.1 & 2.3 & 3 & [\ion{O}{ii}], H$\beta$, [\ion{O}{iii}] \\
625 & 376.6 & 305.8 & 0.3900 & B & 21.41 & 1.47 & - & - & - & - & - & 1.4 & 11.0 & 2 & H/K, H$\delta$, H$\gamma$, G, H$\beta$(ab) \\
103 & -323.9 & 158.1 & 0.3963 & D & 21.88 & 1.69 & - & - & - & - & - & 1.7 & 3.6 & 3 & br \\
540 & 206.2 & 588.7 & 0.4907 & A & 21.03 & 2.48 & $12\pm$2 & - & - & - & - & 1.7 & 10.4 & 4 & G, H$\beta$(ab), br, \ion{Mg}{i} \\
378 & 3.9 & -5.6 & 0.3973 & A & 22.58 & --- & $78\pm$5 & 48 & - & 32 & - & 1.1 & 5.9 & 3 & [\ion{O}{ii}], [\ion{O}{iii}], H$\beta$, H$\gamma$(em), H$\delta$(em)\dots\\
\noalign{\smallskip}\hline\noalign{\smallskip}
209 & -133.6 & 596.3 & 0.4892 & A & 22.48 & 1.20 & $33\pm$2 & 6 & - & 10 & - & 1.4 & 11.7 & 3 & H$\theta$, H$\eta$, H$\zeta$, H$\beta$, [\ion{O}{ii}]\dots\\
315 & -34.2 & 559.5 & 0.4928 & A & 22.28 & 1.68 & $8\pm$1 & - & - & 4 & - & 1.5 & 23.2 & 3 & H/K, H$\zeta$, H$\eta$, H$\theta$, H$\beta$\dots\\
519 & 175.1 & 558.2 & 0.4912 & A & 21.32 & 2.35 & $10\pm$2 & - & - & - & - & 1.8 & 8.1 & 4 & br, G, H$\theta$, H$\eta$, NaD\dots\\
537 & 199.7 & 599.4 & 0.4951 & A & 21.37 & 2.39 & - & - & - & - & - & - & 5.5 & 4 & H/K, H$\eta$, H$\theta$, G \\
546 & 216.7 & 589.7 & 0.4877 & C & 21.81 & 2.33 & - & - & - & - & - & 1.4 & 6.3 & 4 & H, H$\gamma$, H$\eta$ \\
562 & 240.2 & 531.6 & 0.4917 & C & 20.23 & 1.24 & - & - & - & - & $-4\pm$2 & 1.2 & 6.4 & 4 & H$\gamma$, H$\delta$, G \\
567 & 251.0 & 545.7 & 0.4980 & A & 20.30 & 1.58 & $10\pm$2 & - & - & - & - & 1.1 & 8.7 & 4 & H/K, H$\zeta$, H$\eta$, H$\theta$, G\dots\\
\noalign{\smallskip}\hline\noalign{\smallskip}
513 & 169.4 & 393.7 & 0.5016 & C & 22.98 & 1.87 & $21\pm$4 & - & - & - & - & 1.2 & 6.2 & 2 & G, H$\gamma$, H$\delta$, H/K, H$\zeta$ \\
528 & 184.6 & 374.3 & 0.4969 & A & 22.24 & 2.12 & $1\pm$2 & - & - & - & $-5\pm$1 & 1.4 & 9.5 & 2 & H$\delta$, G, H$\gamma$, H$\beta$(ab), H$\theta$\dots\\
556 & 230.6 & 271.1 & 0.4955 & A & 23.47 & 2.35 & $-3\pm$2 & - & - & - & - & 1.7 & 5.0 & 2 & H/K, H$\delta$, G, H$\gamma$, H$\theta$\dots\\
579 & 268.7 & 353.6 & 0.5000 & B & 23.57 & 2.04 & $14\pm$4 & - & - & - & - & 1.5 & 5.9 & 2 & H$\delta$, H/K, H$\theta$, H$\zeta$, G \\
580 & 271.4 & 363.7 & 0.4938 & A & 21.94 & 2.28 & $6\pm$2 & - & - & - & - & 1.5 & 9.3 & 2, 2 & H/K, H$\zeta$, H$\eta$, G, H$\beta$(e+a)\dots\\
610 & 332.6 & 387.9 & 0.4939 & B & 22.45 & 2.16 & $20\pm$3 & - & - & - & - & 1.4 & 5.1 & 2 & br, [\ion{O}{ii}] \\
\noalign{\smallskip}\hline\noalign{\smallskip}
518 & 175.0 & 255.1 & 0.0000 & S & 21.78 & 3.07 & - & - & - & - & - & 0.0 & - & 2 & NaD, \ion{Mg}{i}, bands \\
542 & 209.8 & 261.6 & 0.2981 & A & 22.34 & 0.97 & $31\pm$11 & 9 & 12 & - & - & 1.1 & 9.4 & 2 & [\ion{O}{iii}], [\ion{O}{ii}] \\
544 & 211.3 & 243.4 & 0.1755 & B & 19.98 & 1.39 & - & - & - & - & - & - & 4.7 & 2 & \ion{Mg}{i}, \ion{Fe}{i}, G, H$\gamma$, br\dots\\
552 & 226.9 & 245.0 & 0.3935 & A & 23.02 & 1.09 & $53\pm$8 & 19 & - & - & $-5\pm$2 & 1.1 & 4.8 & 2 & [\ion{O}{ii}], H$\beta$, [\ion{O}{iii}], G, H$\gamma$(em)\dots\\
563 & 240.6 & 256.7 & 0.3797 & A & 20.56 & 1.61 & $8\pm$1 & - & - & 3 & $-3\pm$1 & 1.4 & 21.7 & 2, 2 & H/K, H$\zeta$, H$\eta$, H$\theta$, [\ion{O}{ii}]\dots\\
566 & 246.2 & 279.9 & 0.4000 & A & 21.49 & 1.63 & - & - & - & - & $-5\pm$1 & 1.4 & 10.2 & 2 & H/K, H$\delta$, G, H$\gamma$, H$\beta$(ab)\dots\\
576 & 262.0 & 260.8 & 0.3911 & A & 21.86 & 1.77 & - & - & - & - & - & 1.6 & 12.4 & 2 & br, H/K, H$\theta$, H$\zeta$, H$\eta$\dots\\
\dots & \dots & \dots & \dots & \dots & \dots & \dots & \dots & \dots & \dots & \dots & \dots & \dots & \dots & \dots & \dots \\
\noalign{\smallskip}\hline
\end{tabular}